\documentclass[twocolumn,twocolappendix]{aastex631}

\usepackage[version=4]{mhchem}
\usepackage{graphicx}	
\usepackage{amsmath}	
\usepackage{physics}
\usepackage{enumerate}
\usepackage{CJK}
\usepackage{xcolor}

\graphicspath{{./}{figures/}}

\newcommand{\E}{E_{r}}
\newcommand{\F}{{F}_{r}}
\newcommand{\G}{\mathbb{G}}
\newcommand{\kp}{\kappa_{\rm{P}}}
\newcommand{\kr}{\kappa_{\rm{R}}}
\newcommand{\eg}{e_\text{g}}
\newcommand{\tg}{T_\text{g}}

\newcommand{\M}{M_{\star}}
\newcommand{\rsun}{R$_{\odot}$}

\newcommand{\gcmc}{g$\cdot$cm$^{-3}$}

\newcommand{\msunyr}{M$_{\odot}\cdot$yr$^{-1}$}

\newcommand{\ergs}{erg$\cdot$s$^{-1}$}
\newcommand{\kms}{km$\cdot$s$^{-1}$}

\begin{document}
\begin{CJK*}{UTF8}{gbsn}
\title[Bridging the gap between LRNe and CEE]{Bridging the gap between luminous red novae and common envelope evolution: the role of recombination energy and radiation force}

\correspondingauthor{Zhuo Chen (陈卓)}
\email{chenzhuo\_astro@tsinghua.edu.cn}
\author[0000-0001-7420-9606]{Zhuo Chen (陈卓)}
\affiliation{Institute of Advanced Study, Tsinghua University, Beijing 100084, China}
\affiliation{Department of Astronomy, Tsinghua University, Beijing 100084, China}

\author[0000-0001-6251-5315]{Natalia Ivanova}
\affiliation{Department of Physics, University of Alberta, Edmonton, AB T6G 2E1, Canada}



\begin{abstract}
Luminous red novae (LRNe) and their connection to common envelope evolution (CEE) remain elusive in astrophysics. Here, we present a radiation hydrodynamic model capable of simulating the light curves of material ejected during a CEE. For the first time, the radiation hydrodynamic model incorporates complete recombination physics for hydrogen and helium.  The radiation hydrodynamic equations are solved with {\tt Guangqi}. With time-independent ejecta simulations, we show that the peaks in the light curves are attributed to radiation-dominated ejecta, while the extended plateaus are produced by matter-dominated ejecta. To showcase our model's capability, we fit the light curve of AT2019zhd. The central mass object of $6M_{\odot}$ is assumed based on observations and scaling relations. Our model demonstrates that the ejecta mass of AT2019zhd falls within the range of $0.04M_{\odot}$ to $0.1M_{\odot}$. Additionally, we demonstrate that recombination energy and radiation force acceleration significantly impact the light curves, whereas dust formation has a limited effect during the peak and plateau phases.
\end{abstract}

\keywords{\href{http://astrothesaurus.org/uat/1851}{Transient sources(1851)} --- \href{http://astrothesaurus.org/uat/2154}{Common envelope evolution(2154)}}

\section{Introduction}

Since the discovery of V1309 Sco \citep{tylenda2011}, a luminous red nova (LRN), also a confirmed binary merger, many more new LRNe have been discovered \citep{kurtenkov2015,blagorodnova2017,cai2019,pastorello2019,blagorodnova2021,pastorello2021a,pastorello2021b,cai2022a,cai2022b}, to name a few. It is argued that, at least for some of the LRNe, the progenitors are binary stars that have undergone common envelope evolution (CEE) \citep{2013Sci...339..433I} - one of the most mysterious events in binary evolution. Bridging the gap between the observables of LRNe and theoretical models of CEE could be crucial to getting a better understanding of binary evolution \citep{chen2024}, including cataclysmic variables \citep{warner2003}, X-ray binaries \citep{reig2011}, gravitational wave sources \citep{renzo2021}, and type Ia supernovae \citep{wang2012,liu2023}.

Observationally, LRNe exhibit three prominent characteristic features. Firstly, they all have extended plateaus with decreasing effective temperatures in the light curves. This phenomenon suggests a substantial mass ejection from the central object, which cools as it expands outward. During this phase, hydrogen and helium may recombine and release their latent heat, contributing to the thermal energy of the plateau \citep{ivanova2013,ivanova2015,matsumoto2022}. Secondly, a notable number of LRNe present a peak preceding the plateau in their light curves. Modeling this peak is challenging due to its exponential increasing and decreasing phases. \citet{pejcha2017} employed a hydrodynamic model with post-processing to fit the slowly increasing luminosity before the peak, attributing the peak to an eruption-like event. Although \citet{matsumoto2022} successfully fitted the rapidly decreasing luminosity after the peak with their hot ejecta model in 1D, the rapidly increasing phase remains a puzzle. Thirdly, LRNe frequently exhibit strong H$\alpha$ emission \citep{munari2002,stritzinger2020,blagorodnova2021,pastorello2021a}, indicating a possible collision between the ejecta and circumstellar matter \citep{blagorodnova2020}.

A prevailing theory on the origin of LRNe implicates the progenitor binary undergoes CEE and experiences a plunge-in phase accompanied with ejection of some envelope material.
Numerous 3D simulations have been conducted to study the CEE dynamics \citep{nandez2014,nandez2015,ivanova2016,nandez2016,ohlmann2016,chamandy2018,iaconi2019,prust2019,reichardt2019,sand2020,gonzalez2023,ropke2023,chamandy2024}.
However, due to the complexity of physical processes and the multi-scale nature of the CEE problem, there is still no consensus on how much mass is ejected during a CEE, even if 3D simulations start with the same initial conditions.
Furthermore, predicting the observational appearance of these simulations is inherently difficult \citep{hatful2021}. Notably, radiation hydrodynamics, a crucial factor in CEE, is often absent from these 3D models due to its difficulty and is easier to consider in 1D models \citep{soker2018,bronner2023,oconnor2023}. Recent work by \citet{matsumoto2022} demonstrates the effectiveness of a cooling shell model, devoid of radiation transport and radiation hydrodynamics, in producing LRNe-like light curves.

However, there is still a lack of first principle models that can relate CEE to LRNe. In CEE, the plunge-in phase marks the rapid conversion of the gravitational potential energy into the kinetic energy. Some kinetic energy would be converted to thermal energy via a shock between the plunge-in star and the envelope \citep{macleod2017}. We would expect the shocked gas to be close to adiabatic because of the high optical depth inside a common envelope (CE). The shocked gas may be accelerated in the radial direction due to high radiation pressure and radiation force acceleration -- this has been overlooked previously -- and become an eruption-like event, eventually appearing as an LRN. CEE is a 3D problem by nature. However, solving 3D radiation hydrodynamic equations with complex physics is computationally demanding. Therefore, as the first step, and hoping to resolve the microphysics better, we start by approximating the radiation hydrodynamic problem with 1D spherical symmetry. In a follow-up work, we will revisit this problem with 2D axisymmetric models.

This Letter introduces the first radiation hydrodynamic model that comprehensively incorporates recombination energies of \ce{H} and \ce{He}, radiation transport, and radiation force acceleration. The model can successfully produce 
both the peak and plateau phases of the light curve, allowing for the estimation of ejecta mass through curve fitting. 
The organization of the Letter is as follows: Section \ref{sec:method} presents the physical model, governing equations, initial and boundary conditions, and numerical setups, including Adaptive Mesh Refinement (AMR) criteria. Simulation results are presented in Section \ref{sec:results}. The Letter concludes in Section \ref{sec:con}.

\section{Methods}\label{sec:method}

\subsection{Physical model}\label{sec:physics}

We adopt the following radiation hydrodynamic equations to model the evolution of the LRNe in a 1D spherical coordinate,

\begin{eqnarray}
    \pdv{\rho}{t}+\frac{1}{r^2}\pdv{r}(r^{2}\rho v)&=&0,   \label{eqn:mass}\\
    \pdv{\rho v}{t}+\frac{1}{r^2}\pdv{r}(r^{2}\rho v^{2})&=&-\pdv{p}{r}+\rho(a_{\rm{rad}}-g),   \label{eqn:mom}\\
    \pdv{E}{t}+\frac{1}{r^2}\pdv{r}[r^{2}(E+p)v]&=&\rho v(a_{\rm{rad}}-g)+\G,   \label{eqn:hydroenergy}\\
    \pdv{\E}{t}+\frac{1}{r^2}\pdv{r}[r^{2}(\F+\E v)]&=&-\G,    \label{eqn:radenergy}
\end{eqnarray}
where $\rho,v,p$, and $E$ are the density, radial velocity, pressure, and total energy of the gas. In addition, $a_{\rm{rad}}$ (to be explained later) and $g=G\M/r^{2}$ are the radiation and gravitational force acceleration, respectively, where $G$ is the gravitational constant and $\M$ is the mass of the central object. The time and coordinate are denoted by $t$ and $r$. The radiation-related variables $\E$, $\F$, and $\G$ are explained later in this section.

We adopt a simple hydrogen and helium mixture equation of state (EoS, see Appendix \ref{app:eos}),

\begin{eqnarray}
    E&=&\rho\frac{v^{2}}{2}+\eg(\rho,\tg)    \\
    p&=&\sum_{i}n_{i}k_{b}\tg   
\end{eqnarray}
where $i$ is the species index, $n_{i}$ is the number density of species $i$, $\eg$ and $\tg$ are the internal energy and temperature of the gas, $k_{b}$ is the Boltzmann constant. Throughout this Letter, we assume the hydrogen mass ratio $X=0.74$ and the helium mass ratio $Y=0.26$ for simplicity, because metal does not contribute much to gas thermodynamics. The hydrodynamics, together with the complex EoS, is solved by the approximate HLLC Riemann solver in \citet{chen2019}.

To solve the radiation transport problem, we use the flux-limited diffusion (FLD) approximation, which relates the radiation flux $\F$ to the gradient of radiation energy $\partial\E/\partial r$ with a flux limiter $\lambda$ \citep{levermore1981}, i.e.,
\begin{eqnarray}
    \F&=&-\frac{c\lambda(R)}{\kr\rho}\pdv{\E}{r},    \\
    \lambda(R)&=&\frac{2+R}{6+3R+R^2},   \\
    R&=&\frac{|\partial\E/\partial r|}{\kr\rho\E},
\end{eqnarray}
where $\kr$ is the Rosseland mean opacity (see Appendix \ref{app:opacity}). The flux limiter has the property that 
\begin{equation}\label{eqn:asymptotic}
\frac{c\lambda}{\kr\rho}\pdv{\E}{r}\rightarrow\begin{cases}
	\frac{c}{3\kr\rho}\pdv{\E}{r}&\quad\quad\rm{optically\ thick,	}\\
	c\E &\quad\quad\rm{optically\ thin.}\\
\end{cases}
\end{equation}

The radiation and gas energy coupling is modeled by solving the following equations implicitly with sub-timesteps,
\begin{eqnarray}
    \pdv{\eg}{t}&=&\mathbb{G}, \\
    \pdv{\E}{t}+\frac{1}{r^2}\pdv{r}[r^{2}(\F+\E v)]&=&-\mathbb{G},
\end{eqnarray}
where $\mathbb{G}=\kp\rho c(\E-a_{\rm{r}}\tg^{4})$ is the energy coupling strength, $\kp$ is the Planck mean opacity, and $a_{\rm{r}}$ is the radiation constant. Meanwhile, the radiation and gravitational acceleration are integrated explicitly through,
\begin{equation}
    \pdv{\rho v}{t}=\rho(a_{\rm{rad}}-g),
\end{equation}
where $a_{\rm{rad}}=\kr\F/c$ is the radiation force acceleration and $c$ is the speed of the light.

We use {\tt Guangqi}\citep{chen2024b} to solve Equation \ref{eqn:mass}-\ref{eqn:radenergy}. {\tt Guangqi} is a second-order in time and space accurate and finite volume radiation hydrodynamic code. It has HLLC Riemann solvers, realistic EoS \citep{chen2019}, and adaptive mesh refinement (AMR). {\tt Guangqi} solves the radiation transport problem with FLD approximation implicitly (similar to \citet{kolb2013}), using iterative solvers from {\tt Petsc} \citep{petsc-efficient,petsc-user-ref}. Currently, {\tt Guangqi} has spherical and Cartesian geometry in 1D and 2D.

\subsection{Initial and boundary conditions}\label{sec:ibc}

We set our initial condition to be an outflow at escape velocity with a constant mass loss rate,
\begin{eqnarray}
    \rho_{\rm{init}}(r)&=&\rho_{0}(r_{\rm{in}}/r)^{3/2}, \\
    v_{\rm{init}}(r)&=&\sqrt{2G\M/r}   \\
    T_{\rm{init}}(r)&=&T_{0}(r_{\rm{in}}/r)
\end{eqnarray}
where $\rho_{0}=10^{-13}$\gcmc, $T_{0}=1000$K, and $r_{\rm{in}}=10\text{R}_{\odot}$ is the inner radius of the computational domain. The constant mass loss rate is $\dot{M}=2.30\times10^{-7}$\msunyr, which is low compared to the mass loss rate of the ejecta of the CEE. Radiation's initial condition is assumed to be in local thermal equilibrium (LTE) with gas. The initial mass and energy in the computational domain are $8.52\times10^{-7}\text{M}_{\odot}$ and $3.68\times10^{35}$erg, respectively. They are significantly smaller than the ejecta's mass and energy, and our light curve results are insensitive to the initial condition.

The outer boundary is free, i.e., the gas and radiation can leave the computational domain freely. In our simulations, we set the out boundary at $r_{\rm{out}}=$4000\rsun. The free boundary for the radiation is,
\begin{eqnarray}\label{eqn:radouter}
    \pdv{(r^{2}\E)}{r}=0.
\end{eqnarray}
This outer boundary condition means that the radiation flux is optically thin and outward. We can confirm that the outer region of our computational domain is indeed optically thin (see Section \ref{sec:at2019zhd}) and always outward. We calculated the luminosity at the outer boundary by,
\begin{equation}
    L=4\pi r_{\rm{out}}^{2}\F(r_{\rm{out}})
\end{equation}

The inner boundary is time-dependent. We specify the time-dependent density, velocity, and temperature of the ejecta at the inner boundary, i.e., $[\rho(t),v_{\rm{ej}}(t),\tg(t)]$.
\begin{eqnarray}
    v_{\rm{ej}}&=&f_{\rm ej}(t) \sqrt{2G\M/r_{\rm{in}}},   \label{eqn:singlev}\\
    \rho&=&\dot{M}/(4\pi r_{\rm{in}}^{2}v_{\rm{ej}}), \label{eqn:singlerho}\\
    \tg&=&\alpha m_{\ce{H}}v_{\rm{ej}}^{2}/2k_{b}, \label{eqn:singlet}
\end{eqnarray}
where $f_{\rm ej}(t)$ can be a time-dependent factor, $\dot{M}$ is the mass loss rate, and we introduce $\alpha$ to characterize the temperature of the ejecta. When the ejection of the CEE stops, the inner boundary of the gas is changed to free to let the fallback gas pass through the inner boundary, therefore, we do not model any fallback shocks and fallback accretion disks. Meanwhile, radiation transport is turned off at the inner boundary. Mathematically, we use the zero-gradient ($\partial\E/\partial r=0$) radiation inner boundary condition.

We assume the gas and radiation are in LTE at the inner boundary because the density and opacity are very high inside the CE, and the thermal timescale is short.

\subsection{Numerical setups}\label{sec:setups}

We adopt a uniform base grid with spherical geometry. The computational domain is $r\in[10,4000]$\rsun, and the base resolution is $N=1536$. We add 5 levels of static mesh refinement where $r\in[10,15]$\rsun to resolve the strong gradient of the gravitational potential. Each level of mesh refinement doubles the resolution. We also adaptively refine zones with temperature gradients up to 5 levels. The mesh refine and derefine criterion are as follows,
\begin{eqnarray}
    \text{refine}\quad\frac{|T_{g,i+1}-T_{g,i-1}|}{2T_{g,i}}&>&0.03, \\
    \text{derefine}\quad\frac{|T_{g,i+1}-T_{g,i-1}|}{2T_{g,i}}&<&0.001,
\end{eqnarray}
where $i\pm1$ in the subscript represents the cell index (not to be confused with species index). We can capture the shock, radiative layer, and radiation-dominated zones with photon trapping with AMR. The finest cell has a length of $8.12\times10^{-2}$\rsun. The Courant--Friedrichs--Lewy number is 0.95 in our simulations. The simulation time is $4.5\times10^{6}$s=52.08 days.

\section{Results}\label{sec:results}

We first show some simple simulations and get a sense of the correspondence between the properties of the ejecta and the light curves. After that, we fit the light curve of AT2019zhd with a more complex ejecta.

\subsection{Simple ejecta}

For simplicity, in this subsection, we set $\M=6$M$_{\odot}$, $r_{\rm{in}}=10$\rsun. We consider seven models and list the physical properties of the seven models in Table \ref{tab:simplemodels}. In what follows, we refer to a model as radiation-dominated if $\E/\eg\gg1$, and as matter-dominated if $\E/\eg<1$. In particular, we calculate the total energy $E_{\rm{total}}$ of the ejecta by,
\begin{equation}
    E_{\rm{total}}=4\pi r_{\rm{in}}^{2}(\E+\eg)f_{\rm{ej}}\sqrt{2G\M/r_{\rm{in}}}t_{\rm{ej}},
\end{equation}
where $t_{\rm{ej}}$ is the duration of the ejection.
\begin{table*}
    \centering
    \begin{tabular}{cccccccccc}\hline
    model&  $f_{\rm{ej}}$   &   $t_{\rm{ej}}$   & $\dot{M}$ &   $\Delta M$   &   $\rho$   &   $\alpha$    &   $\tg$   &   $\E/\eg$  & $E_{\rm{total}}$    \\
         &  &   [days]   &   [$\text{M}_{\odot}\cdot\text{yr}^{-1}$]  &   [M$_{\odot}$]   &   [\gcmc]   &               &   [K]     & &    [erg]          \\\hline
    m01a18  &   1   &   1     &    0.1     &  $2.74\times10^{-4}$&   $2.17\times10^{-8}$     &   0.018       &   $2.50\times10^{5}$ & 19.9  &   $7.76\times10^{44}$   \\
    m02a045v2  &   2   &   0.5     &    0.2     &  $2.74\times10^{-4}$&   $2.17\times10^{-8}$    &   0.0045       &   $2.50\times10^{5}$ & 19.9 &  $7.76\times10^{44}$   \\
    m01a09  &   1   &   1     &    0.1     &  $2.74\times10^{-4}$   &   $2.17\times10^{-8}$  &   0.009       &   $1.25\times10^{5}$ & 2.01  &   $6.91\times10^{43}$  \\
    m05a18  &   1   &   1     &    0.5     &  $1.37\times10^{-3}$   &   $1.08\times10^{-7}$  &   0.018       &   $2.50\times10^{5}$ & 3.98  &   $9.24\times10^{44}$  \\
    m05a09  &   1   &   1     &    0.5    &  $1.37\times10^{-3}$   &   $1.08\times10^{-7}$  &   0.009       &   $1.25\times10^{5}$ & 0.40  &   $1.61\times10^{43}$  \\
    m25a18  &   1   &   1     &    2.5    &  $6.85\times10^{-3}$   &   $5.42\times10^{-7}$  &   0.018       &   $2.50\times10^{5}$ & 0.80  &   $1.66\times10^{45}$  \\
    m25a09  &   1   &   1     &    2.5     &  $6.85\times10^{-3}$   &   $5.42\times10^{-7}$  &   0.009       &   $1.25\times10^{5}$ & 0.08  &   $6.20\times10^{44}$   \\\hline
    \end{tabular}
    \caption{From left to right, model name, ejecta's velocity factor, ejecta duration $t_{\rm{ej}}$, mass loss rate $\dot{M}$, cumulative mass loss $\Delta M$, ejecta's density, $\alpha$, the temperature of the ejecta, the ratio of $\E$ and $\eg$, and the total energy of the ejecta of the seven simple models.}
    \label{tab:simplemodels}
\end{table*}

Figure \ref{tab:simplemodels} shows the light curves of the simple models. We can see that the light curves of m02a045v2 and m01a18 have pronounced peaks, i.e., rapid, exponential increases and decreases, while m25a18 and m25a09 have long plateaus. The light curves of other models are in between. Because the physical system is highly nonlinear, we provide a qualitative discussion here. The optical depth of an ejecta decreases as it expands, and the ejecta cools faster as the optical depth decreases. The more mass in an ejecta, the higher the optical depth; the faster the ejecta, the faster the optical depth decreases. Therefore, m25a18 and m25a09 have long plateaus because their ejecta are the most massive ones among the seven simple models; m25a18 has a longer plateau than m25a09 because m25a18 has a higher energy budget and thus a longer cooling timescale. On the other hand, m02a045v2 has a sharper peak than m01a18 because its expansion speed is faster, resulting in a more rapid decrease in the optical depth and a shorter cooling timescale. Because m02a045v2 and m01a18 have the same amount of energy budget, m02a045v2 has a higher luminosity peak.

Overall, we can also relate that we observe the luminosity peak formation (exponential increase and decrease) in the case of radiation-dominating ejecta. In contrast, the plateau is the feature observed in the case of matter-dominated ejecta.

\begin{figure}
    \centering
    \includegraphics[width=\columnwidth]{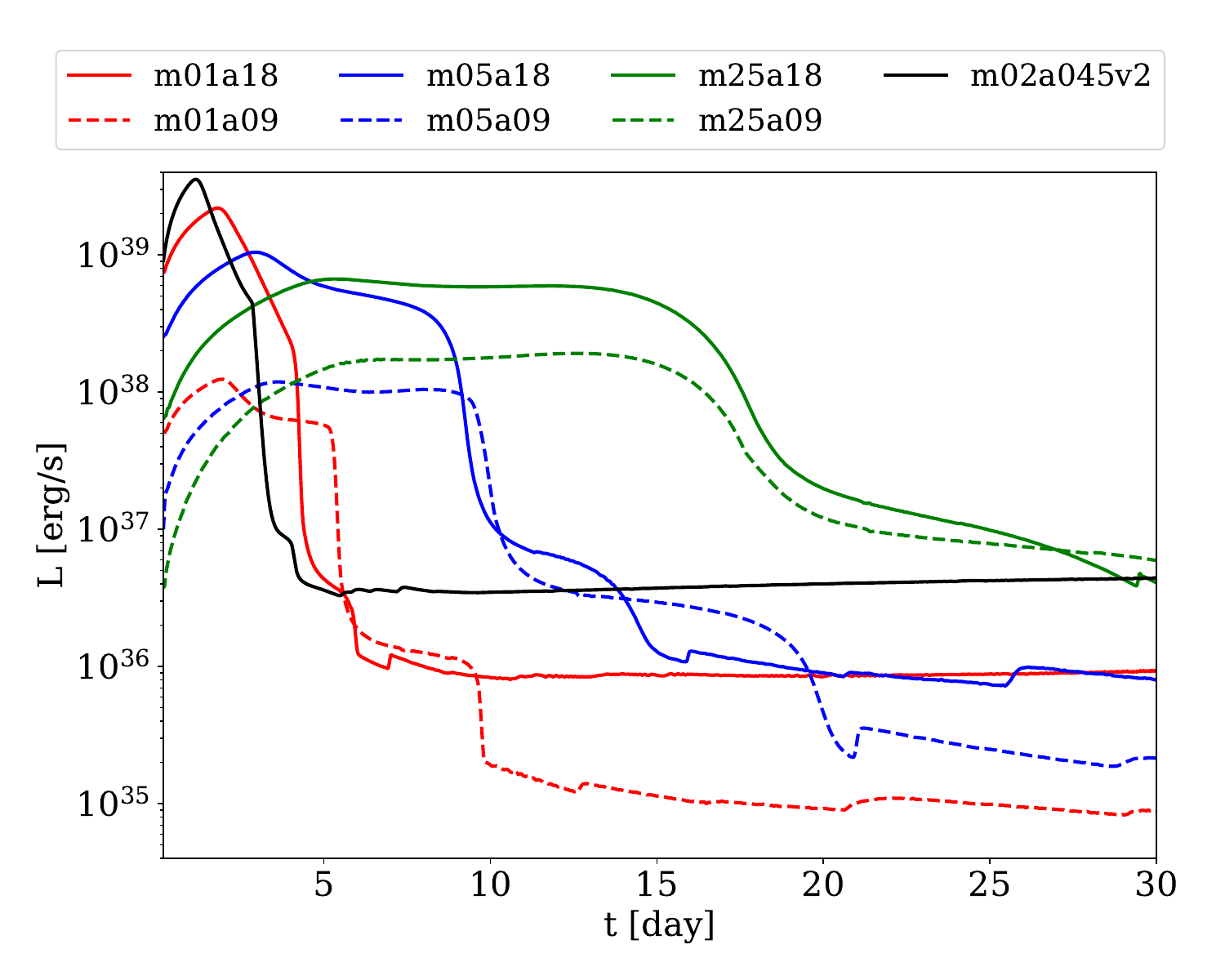}
    \caption{The light curves of simple simulations with different mass and temperature of the ejecta. The physical parameters of the ejecta are listed in Table\ref{tab:simplemodels}.}
    \label{fig:single}
\end{figure}

\subsection{Fitting AT2019zhd}\label{sec:at2019zhd}

We choose AT2019zhd as a fitting example because it is a recently well-observed LRN and its light curve resembles V1309 Sco, whose progenitor is a confirmed binary \citep{tylenda2011,pastorello2021a}. 
The mass of the progenitor of AT2019zhd is unclear \citep{pastorello2021a}. We take the following steps to estimate the total mass of the merger product. 

\begin{enumerate}[1.]
    \item The total mass of V1309 Sco is estimated to be $1-2M_{\odot}$ \citep{tylenda2011}.
    \item The peak luminosity of AT2019zhd is roughly $L_{\rm{AT2019zhd}}=2.08\times10^{39}$\ergs, and the peak captured luminosity of V1309 Sco is $L_{\rm{V1309 Sco}}=1.26\times10^{38}$\ergs. Therefore, $L_{\rm{AT2019zhd}}/L_{\rm{V1309 Sco}}=16.5$.
    \item The total radiation energy released during the peak and the plateau of AT2019zhd is roughly 10 times more than V1309 Sco.
\end{enumerate}
If we assume that the luminosity energy mainly comes from the release of the gravitational potential energy, it may scale as,
\begin{eqnarray}\label{eqn:scaling}
    L\sim M^{2}.
\end{eqnarray}
This scaling relation motivates us to set the mass of the central object to be the same as was adopted for the simple models, $6M_{\odot}$, and we set $\alpha=0.018$ in this subsection.

\begin{figure*}
    \centering
    \includegraphics[width=\textwidth]{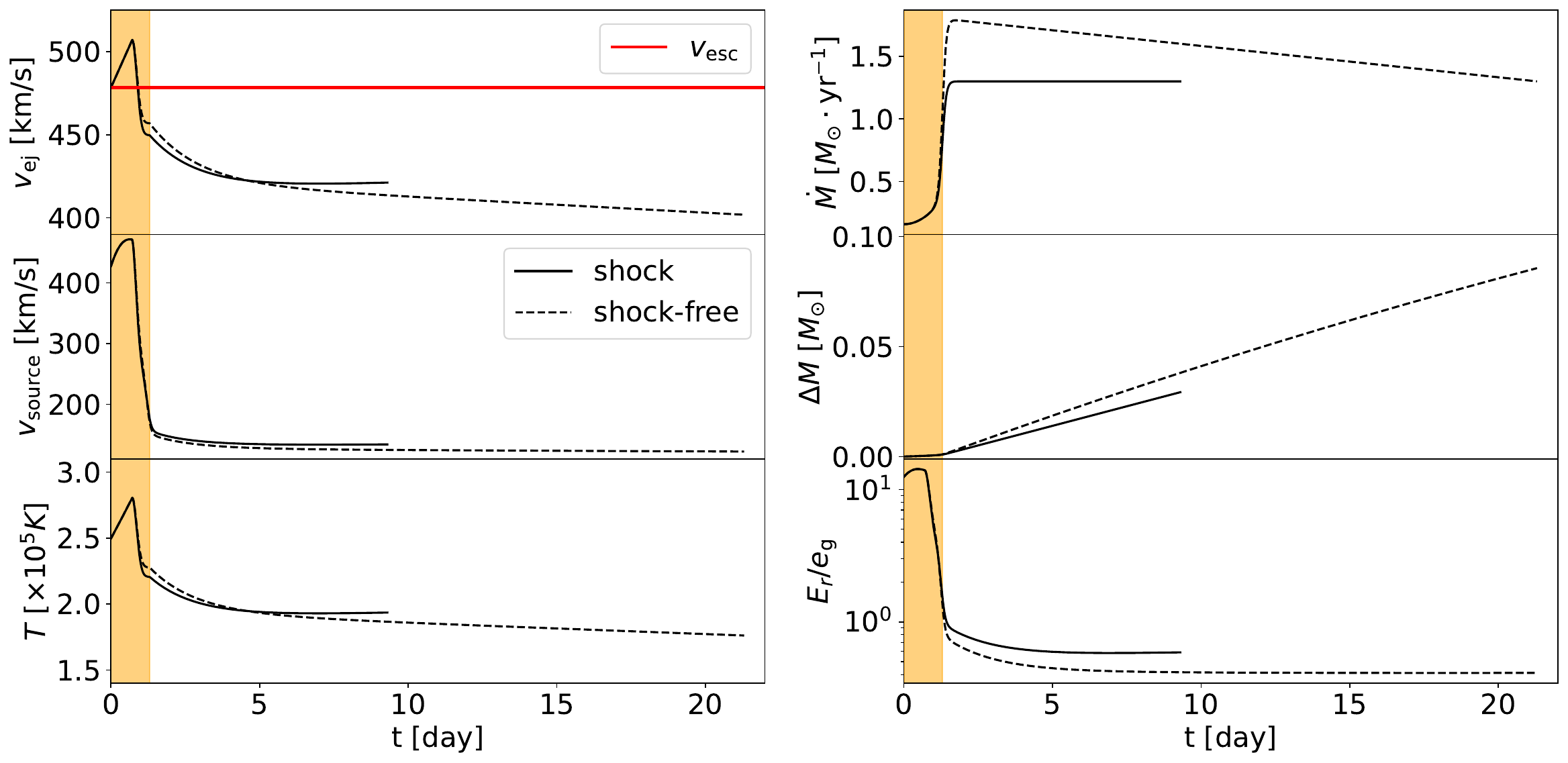}
    \caption{The solid and dashed lines are the time-dependent inner boundary conditions of the shock and shock-free models, respectively. The red line indicates the escape velocity at the inner boundary. The orange region denotes the radiation-dominated ejecta. On the left panel, from top to bottom, each plot shows the velocity of the ejecta, $v_{\rm{source}}$ calculated from Equation \ref{eqn:vsource}, and the temperature of the ejecta. On the right panel, from the top to the bottom, they are the mass loss rate, cumulative mass loss, and $\E/\eg$, respectively. The shock model and shock-free model also differ in the duration of the ejecta.}
    \label{fig:innerbound}
\end{figure*}

Unlike the case of simple models, the inner boundary conditions here are time-dependent, see Figure \ref{fig:innerbound}. As a result,
the time-dependent ejecta is initially radiation-dominated (denoted by the orange color in Figure~\ref{fig:innerbound}) and then transits to matter-dominated. We anticipate that the initial high-temperature ejecta is produced by a shock between the rapidly plunge-in companion star and the envelope \citep{macleod2017}. The plunge-in speed $v_{p}$ is comparable to the Keplerian speed at the plunge-in radius $r_{p}$, but much larger than the envelope's speed $v_{\rm{env}}$. On the other hand, the kinetic energy required to produce $\E$ and $\eg$ in the ejecta, converted to speed, can be calculated by,
\begin{equation}
    v_{\rm{source}}=\sqrt{\frac{2(\E+\eg)}{\rho}},  \label{eqn:vsource}
\end{equation}
where $\rho$ is the ejecta's density. Due to energy conversion, 
\begin{equation}
    v_{p}^{2}\approx\frac{G\M}{r_{p}}>v_{\rm{env}}^{2}+v_{\rm{source}}^{2}.
\end{equation}
The maximum $v_{\rm{source}}$ in Figure \ref{fig:innerbound} is 448\kms, which means that $r_{p}<5.69$\rsun. More sophisticated stellar and binary evolution analysis should be considered to further narrow down $r_{p}$ and $\M$ \citep{ge2010,ge2015,ge2020}.

We carry out simulations of two sub-models: the shock model and the shock-free model. The shock model has an ejecta with a slightly increasing speed during the late stage and the shock-free model has an ejecta with a slightly decreasing speed during the late stage. Figure~\ref{fig:innerbound} shows the time-dependent variables of the two sub-models.  The functions that generate these variables can be found in Appendix \ref{app:boundconds}.

\begin{figure}
    \centering
    \includegraphics[width=\columnwidth]{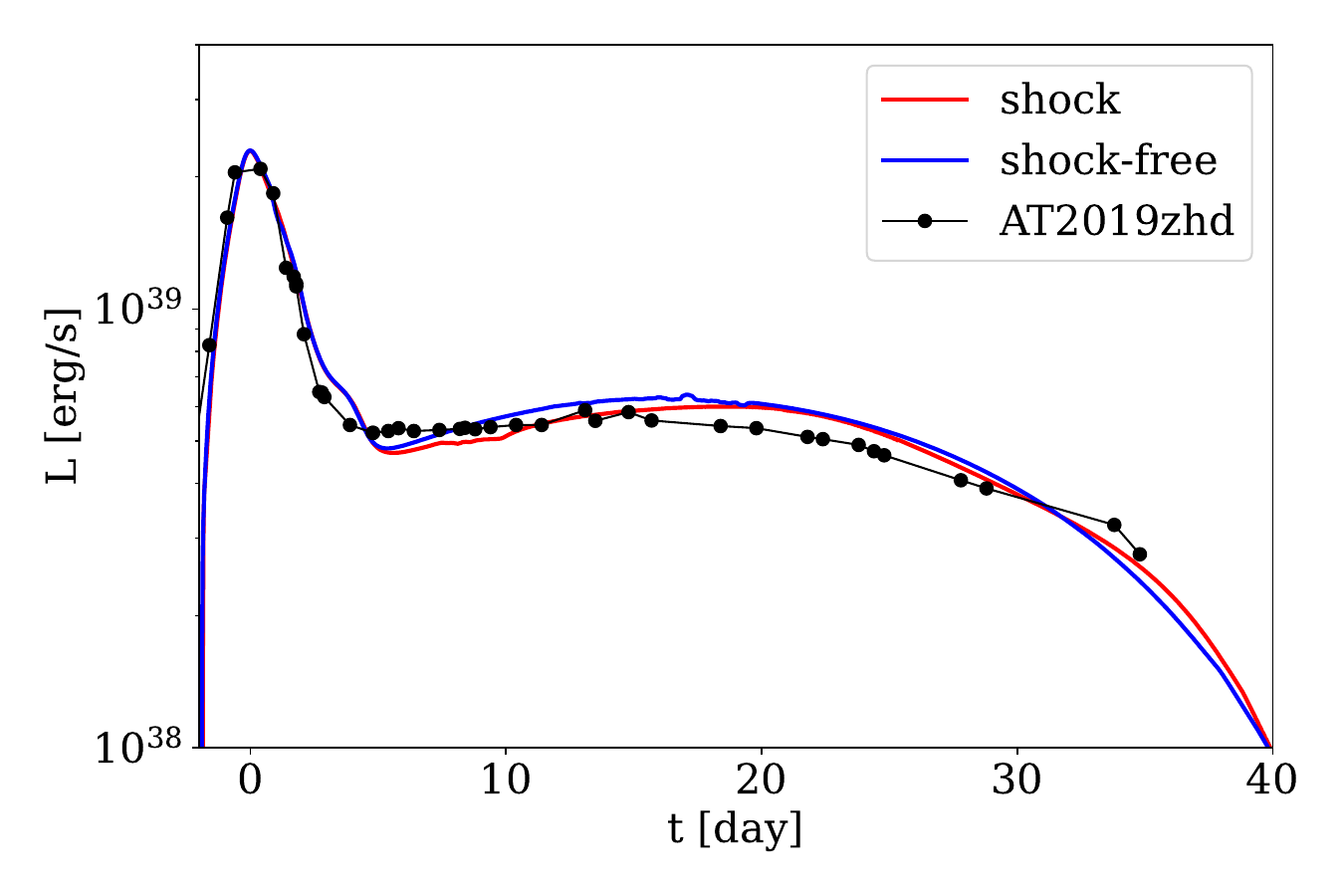}
    \caption{The solid line with black dots is the observed light curve of AT 2019zhd \citep{pastorello2021a}, and the red and blue solid lines are the light curves of the shock and the shock-free models, respectively.}
    \label{fig:at2019zhd}
\end{figure}

\begin{figure*}
    \centering
    \includegraphics[width=\textwidth]{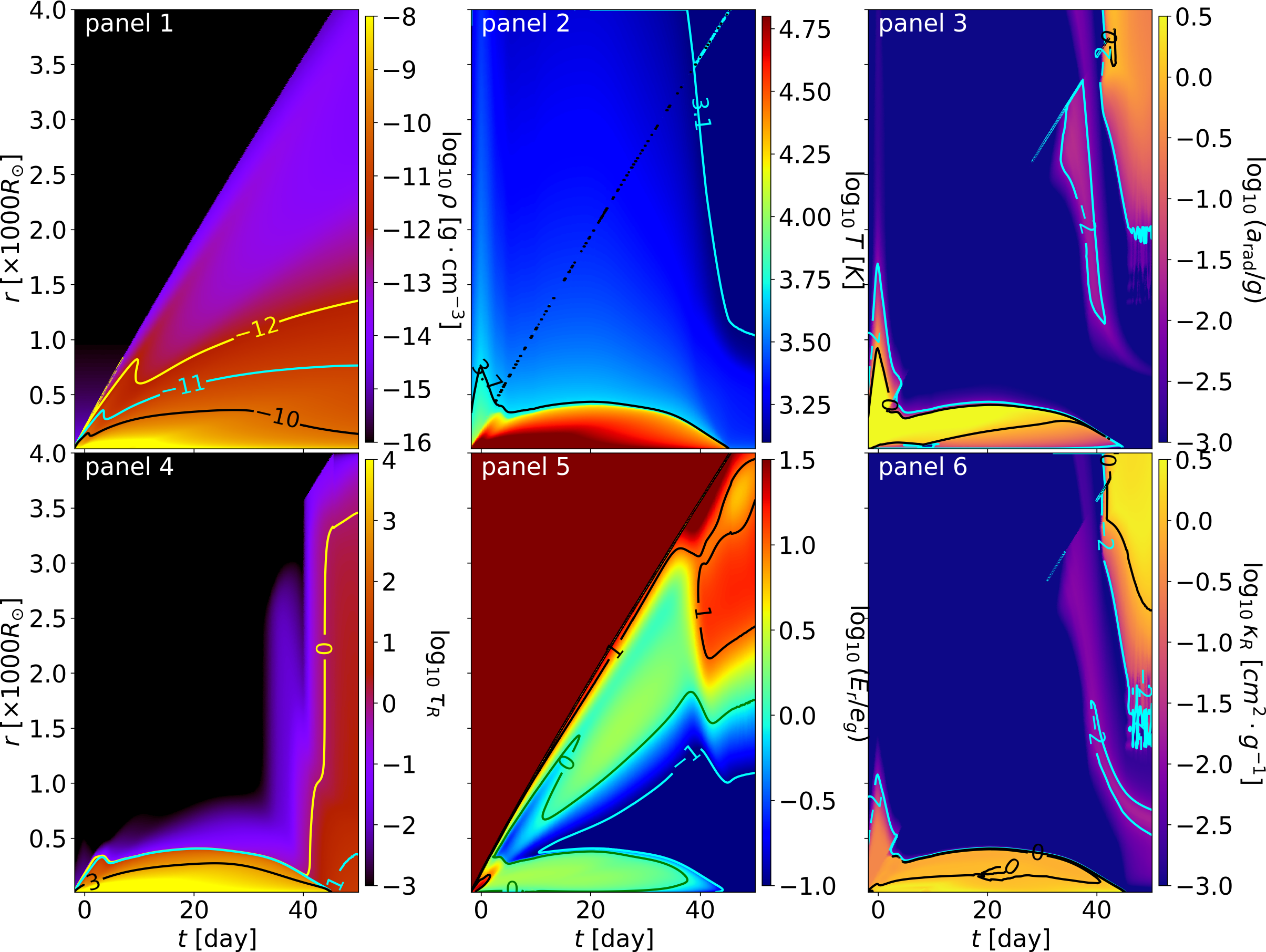}
    \caption{The time evolution of the shock model. In each panel, the x-axis is time in days, and the y-axis is radius in 1000\rsun. From panel 1 to 6, they are $\log_{10}\rho$ with contours of $[-10,-11,-12]$, $\log_{10}\tg$ with contours of $[3.1,3.7]$, and $\log_{10}(a_{\rm{rad}}/g)$ with contours of $[-2,0]$, the optical depth $\log_{10}\tau_{R}$ with contours of $[0,1,3]$, $\log_{10}(\E/\eg)$ with contours of $[-1,0,1]$, and $\log_{10}\kr$ with contours of $[-2,0]$.}
    \label{fig:timeevolmisc}
\end{figure*}

Figure \ref{fig:at2019zhd} shows the light curves of the shock and shock-free models. We have shifted the time axis to let the peak be located at $t=0$. We can see that the light curves of the shock and shock-free models both resemble AT2019zhd. Consequently, the time evolution of the shock-free model is similar to the shock model, therefore, we just show the time evolution of the shock model in Figure \ref{fig:timeevolmisc} to save space. In particular, we calculate the optical depth of the radiation flux by,
\begin{equation}
    \tau_{R}(r)=\int_{r}^{r_{\rm{out}}}\rho\kr dr.
\end{equation}
The time axis is adjusted to be consistent with the light curves. The time evolution of \ce{H} and \ce{He} species can be found in Appendix \ref{app:hhe}, since they are determined by $\rho$ and $\tg$.

There are several prominent results in the time evolution plots, see Figure~5.
\begin{enumerate}[1.]
    \item In panel 2, the temperature profile has a spike near the peak of the luminosity. This is when the radiation-dominated ejecta becomes optically thin and the radiation heats the surrounding gas. 
   
    \item The observationally important ejecta material is the one that has a temperature of about 5000K. The ejecta's matter that is colder than 5000K is closer to the observer but is mostly transparent, revealing behind it the ejecta's matter with a temperature of about 5000K. The time-evolution of the surface $\log_{10}\tg=3.7$ (see panel 2) resembles the shape of the light curve in Figure \ref{fig:at2019zhd}.
    
    \item Panel 3 shows that the $a_{\rm{rad}}$ exceeds the $g$ at small radii during the first 30 days, and at large radii during the late stage. This phenomenon is consistent with the evolution of $\kr$. 
    When $10^4\la \tg\la 10^5$K, most of the gas is ionized and $\kr$ is large, radiation flux can be significant.
    When $\tg<1400$K, dust may form and provide the opacity, and the $a_{\rm{rad}}$ becomes significant again. At this stage, the LRNe is similar to the asymptotic-giant-branch (AGB) stars \citep{hofner2018}.
    
    \item In panel 4, the radius of $\tau_{R}=1$ expands rapidly when the inner region of the ejecta cools off because dust can form at low temperatures and obscure the object.
    
    \item The inner region ($r<2000$\rsun) transits from radiation-dominated to matter-dominated as the ejecta expands into the ambient and cools off. It can be seen in panel 5 of Figure \ref{fig:timeevolmisc}.
    
\end{enumerate}
Our model is comprehensive in the sense that we incorporate many physical processes. To address the impact of each specific physics on the light curve, we run some companion simulations by turning off the corresponding physics in Appendix \ref{sub:impacteos}, \ref{sub:frad}, and \ref{sub:dust}. We show that our calculations are converged in Appendix \ref{app:convergence}.

\section{Conclusions}\label{sec:con}

We present a 1D radiation hydrodynamic model that can obtain light curves of LRNe events. For the first time, our model incorporates recombination physics, radiation transport, and radiation force acceleration of a CEE into consideration. There are several key new physics that we determined to be intrinsic to CEE and LRNe.

First, we find that the light curve's peak and plateau are formed differently. Specifically, the peak is formed primarily by radiation-dominated ($\E/\eg\gg1$), and the plateau is formed by matter-dominated ejecta ($\E/\eg<1$).

Secondly, radiation force plays a significant role in driving the ejecta outwards (panel 3 of Figure \ref{fig:timeevolmisc}). This occurs where opacity is high, specifically when both \ce{H} and \ce{He} are still ionized. Radiation force may help to eject matter that initially has sub-escape velocity.
    
At last, rapid dust formation, assuming it occurs, absorbs radiation and accelerates during the late LRNe stages, resembling environments near AGB stars. While the radiation force helps to accelerate the dusty matter, this has minimal impact on the peak and first plateau of the light curve.

There are several limitations in our model right now. 
For example, the temperature of the ejecta is described by Equation \ref{eqn:singlet}. Thus, we can not strictly distinguish the effect of changing $\M/r_{\rm{in}}$ and $\alpha$. However, one can constrain $\M$ by a scaling relation, for example, Equation \ref{eqn:scaling}, and can constrain $r_{\rm{in}}$ (which is affected by $r_{s}$) with long-term binary evolution models. Moreover, one can further constrain the thermalization efficiency $\alpha$ with realistic 3D radiation hydrodynamic calculations.

The effect of changing initial $\eg$ and  $\E$  in the ejecta is also not fully independent, but radiation transport can, to some extent, help to distinguish the radiation-dominated cases from the matter-dominated cases. 

We also find that shock and shock-free models can produce similar results. In our models, shock and shock-free models refer to whether the speed of the ejecta was increasing or decreasing with time. Indeed, a shock can generate heat and prolong the plateau phase, but adding mass and internal energy to the ejecta can also result in a prolonged plateau in the light curve. We may need other observables, such as H$\alpha$ emission, to distinguish these two scenarios. We can also explore the effect of pre-CEE asymmetric matter distribution around the central binary object; this will be the subject of the follow-up 2D work.

Even with a 1D model, we can fit the AT2019zhd light curve reasonably well. Motivated by observations and scaling relations (Section \ref{sec:at2019zhd}), we adopted the mass of the central object to be $\M=6M_{\odot}$, and show that our choice of $\M$ and $r_{\rm{in}}$ are reasonable. With this choice, we estimate that the mass of the ejecta could be $\Delta M\in[0.04,0.1]M_{\odot}$\footnote{A different choice of the central object mass will lead to different inferred CE ejecta properties if such a fit is found. While our model fits the observations considering reasonable assumptions about ejecta's density, internal and kinetic energies, we do not exclude a possibility of other fits with CE ejecta properties being far outside of considered in this paper.}. As a result, the 1D model that we proposed in this Letter can provide a crucial bridge between CEE theory and the observables of LRNe.

\section*{}
We thank the referee for the careful review and the constructive suggestions that improve the quality of this Letter. The authors thank Andrea Pastorello for sharing AT2019zhd's light curve data with us. Z.C. enjoyed discussing with Xuening Bai (白雪宁), Shude Mao (毛淑德), and Wei Zhu (祝伟). Z.C. is grateful for the support from the National Natural Science Foundation of China (NSFC No.12103028, No.12342501), Tsinghua University Dushi Program, and Shuimu Tsinghua Scholar Program. N.I. acknowledges funding from NSERC Discovery under Grant No. NSERC RGPIN-2019-04277.

\software{{\tt Guangqi 光启} \citep{chen2024b}, {\tt Matplotlib} \citep{hunter2007}, {\tt Petsc} \citep{petsc-efficient,petsc-user-ref}}



\bibliography{lrn}
\bibliographystyle{aasjournal}




\appendix

\section{Equation of state}\label{app:eos}

Realistic EoS can be highly nonlinear. Here, we present a simplified EoS model of \ce{H} and \ce{He} mixture, i.e., the sum of the mass fraction of hydrogen element $X$ and helium element $Y$ equals 1. This model considers the latent heat of hydrogen dissociation and all the ionization transitions of the \ce{H} and \ce{He}. To simplify the calculation, we assume that \ce{H}, \ce{He}, and \ce{e-} have the same temperature, but \ce{H} and \ce{He} have independent thermodynamic subsystems. This thermodynamic system is not as complete as in \citep{tomida2013} but is much easier to solve.

For a thermodynamic system with multi-species, its entropy $s$, pressure $p$, internal energy $\eg$, and heat capacity $C_{V}$ are,
\begin{eqnarray}
    s&=&k_{b}\sum_{i}n_{i}\bigg[1+\frac{d\ln z_{i}}{d\ln\tg}-\ln\bigg(\frac{n_{i}}{z_{i}}\bigg)\bigg],    \\
    p&=&\sum_{i}n_{i}k_{b}\tg, \\
    \eg&=&\sum_{i}n_{i}\xi_{i},    \\
    C_{V}&=&\sum_{i}\pdv{n_{i}}{\tg}\xi_{i}+\sum_{i}n_{i}\pdv{\xi_{i}}{\tg},
\end{eqnarray}
where $i$ is the summation index of all the species, $n_{i}$, $\xi_{i}$, and $z_{i}$ are the number density, particle energy, and partition function of species $i$. The particle energy of hydrogen and helium species, in the regime of the ideal gas law, are described by,
\begin{equation}
    \xi_{i}=3k_{b}\tg/2+\phi_{i},
\end{equation}
where $\phi_{i}$ is each particle's dissociation/ionization binding energy, which is also the source of the latent heat. For example, the ground state of \ce{He} is chosen to be $\phi_{\ce{He}}=0$, and the binding energy of \ce{He^{2+}} is $\phi_{\ce{He^2+}}=24.59+54.42=80.01$eV. The partition functions, not considering the atomic structure, are described by,
\begin{equation}
    z_{i}=\bigg(\frac{2\pi m_{i}k_{b}\tg}{h^2}\bigg)^{3/2}\exp(\frac{-\phi_{i}}{k_{b}\tg}),
\end{equation}
where $m_{i}$ is the mass of each particle species and $h$ is the Planck constant.

\begin{figure}
    \centering
    \includegraphics[width=\columnwidth]{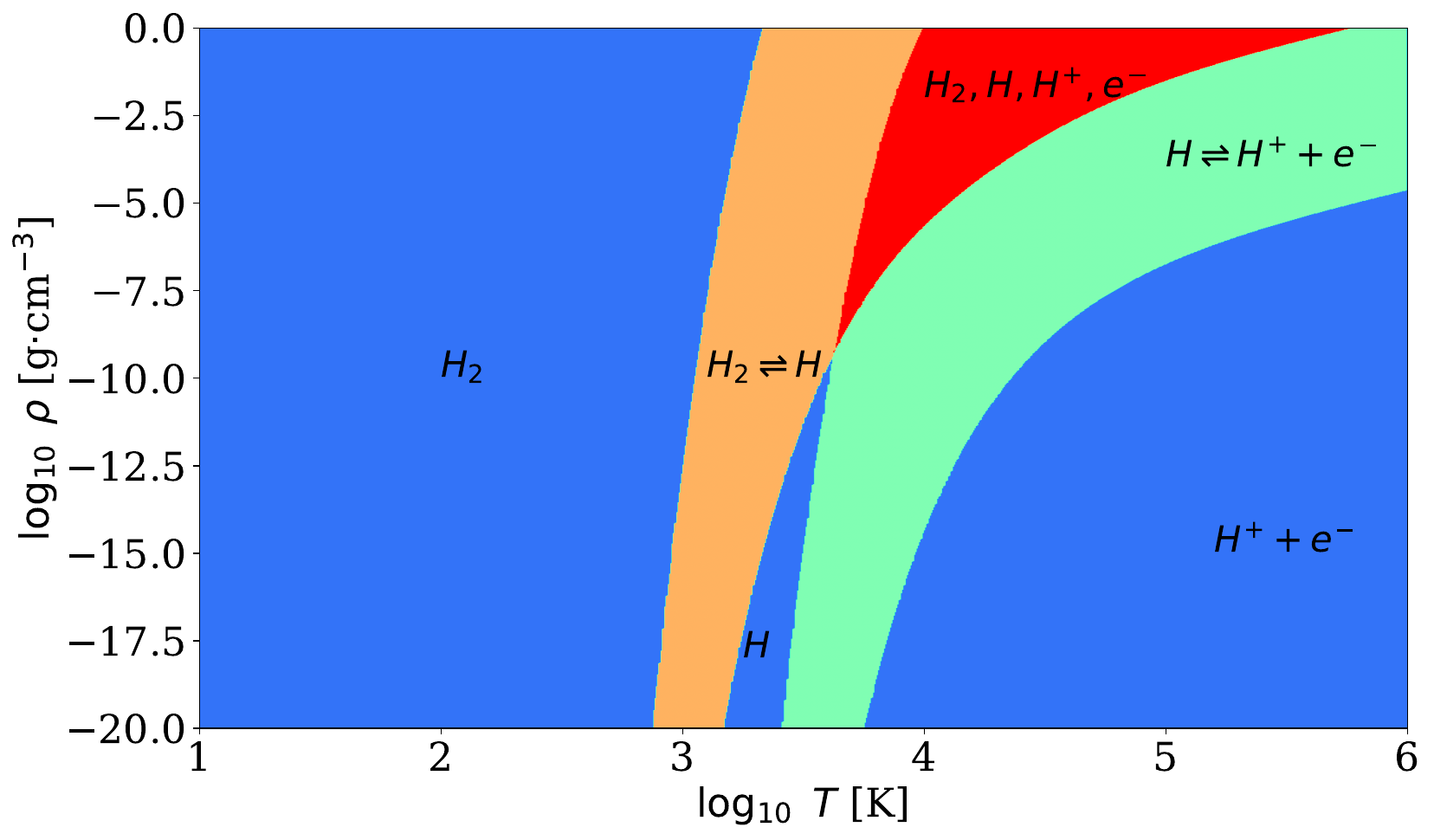}\\
    \includegraphics[width=\columnwidth]{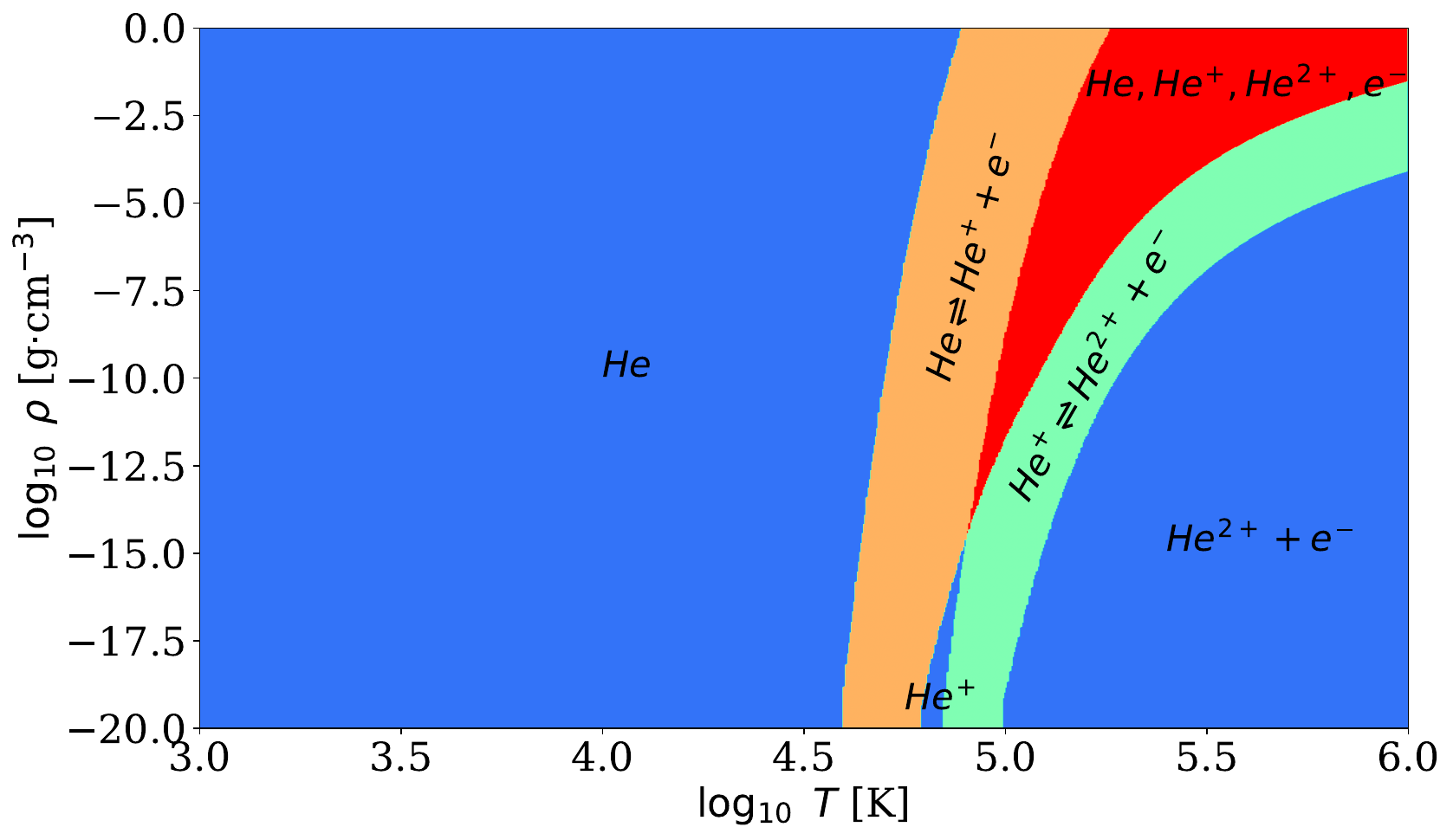}
    \caption{The $\rho$-$T$ diagram of \ce{H} and \ce{He} are divided into 6 zones to improve the computational efficiency. The pure states are colored with blue, and the rest are mixture states that require solving Saha equations.}
    \label{fig:division}
\end{figure}

We also need the adiabatic sound speed $c_s$ and adiabatic index $\gamma$ for the Riemann solver \citep{chen2019}, which is calculated by,
\begin{eqnarray}
    c_{s}&=&\sqrt{\bigg(\pdv{p}{\rho}\bigg)_{T}+\bigg(\pdv{p}{T}\bigg)_{\rho}\bigg(\pdv{T}{\rho}\bigg)_{s}}    \notag\\
    &=&\sqrt{\bigg(\pdv{p}{\rho}\bigg)_{T}-\bigg(\pdv{p}{T}\bigg)_{\rho}\frac{(\partial s/\partial\rho)_{T}}{(\partial s/\partial T)_{\rho}}},  \\
    \gamma&=&\rho c_{s}^{2}/p
\end{eqnarray}
Thus, to calculate $c_s$, we need to know $(\partial s/\partial\rho)_{T}$ and $(\partial s/\partial\rho)_{\rho}$, which means $(\partial n_{i}/\partial\rho)_{T}$ and $(\partial n_{i}/\partial\rho)_{\rho}$. They can be calculated by thermodynamic equilibrium equations, i.e., the Saha equations. To further simplify the EoS model, we divide the $\rho-T$ diagram into pure and mixture states \citep{chen2019}. We define pure states if the mass fraction of a single species (not including electron) exceeds $0.99999$ and all other cases are mixture states. Figure \ref{fig:division} shows the division of \ce{H} and \ce{He}. The blue color represents the pure states of \ce{H} and \ce{He}. In doing so, we can solve fewer Saha equations.

As a first example, let us consider hydrogen's ionization and dissociation reactions (the red color zone in the first panel of Figure \ref{fig:division}).
\begin{eqnarray}
    \ce{H}&\longleftrightarrow&\ce{H+}+\ce{e-},  \\
    \ce{H2}&\longleftrightarrow&2\ce{H}.
\end{eqnarray}

The corresponding Saha equations are
\begin{eqnarray}
    \frac{n_{\ce{H}}^2}{n_{\ce{H2}}}&=&\frac{z_{\ce{H}}^2}{z_{\ce{H2}}}=q_1,  \label{eqn:saha1}\\
    \frac{n_{\ce{H+}}^2}{n_{\ce{H}}}&=&\frac{z_{\ce{H+}}z_{\ce{e-}}}{z_{\ce{H}}}=q_2,   \label{eqn:saha2}
\end{eqnarray}
where we have used the charge neutrality equation
\begin{eqnarray}
    n_{\ce{H+}}=n_{\ce{e-}}
\end{eqnarray}

We also have the number conservation equation,
\begin{eqnarray}\label{eqn:hydronum}
    2n_{\ce{H2}}+n_{\ce{H}}+n_{\ce{H+}}=X\rho/m_{\ce{H}}=\bar{X},
\end{eqnarray}
where $X$ is the mass fraction of $\ce{H}$ of the gas. Substitute the Saha equations into the number conservation equation, and we get a quadratic equation
\begin{eqnarray}\label{eqn:hydrogen_quartic}
    \frac{2n_{\ce{H+}}^4}{q_{1}^2 q_{2}}+\frac{n_{\ce{H+}}^2}{q_{2}}+n_{\ce{H+}}=\bar{X}.
\end{eqnarray}
The solution of Equation \ref{eqn:hydrogen_quartic} is $n_{\ce{H+}}$, and $n_{\ce{H}}$ and $n_{\ce{H2}}$ can be calculated from the Saha equations.

We use the differential form of Equation \ref{eqn:saha1}, \ref{eqn:saha2}, and \ref{eqn:hydronum} to obtain $(\partial n_{i}/\partial\rho)_{T}$ and $(\partial n_{i}/\partial\rho)_{\rho}$ of \ce{H}.
\begin{eqnarray}
\begin{pmatrix}
    -1 & 2 & 0 \\
    0 & -1 & 2 \\
    2n_{\ce{H2}} & n_{\ce{H}} & n_{\ce{H+}}
\end{pmatrix}
\begin{pmatrix}
    d\ln n_{\ce{H}}     \\
    d\ln n_{\ce{H+}}    \\
    d\ln n_{\ce{H2}}
\end{pmatrix}
=
\begin{pmatrix}
    d\ln q_{1}      \\
    d\ln q_{2}      \\
    \bar{X}d\ln \bar{X}
\end{pmatrix}
\end{eqnarray}

If only one reaction is present, the thermodynamic system is trivial to solve.

For \ce{He}, let us consider the two reactions (the red color zone in the second panel of Figure \ref{fig:division})
\begin{eqnarray}
    \ce{He}&\longleftrightarrow&\ce{He+}+\ce{e-},    \\
    \ce{He+}&\longleftrightarrow&\ce{He^{2+}}+\ce{e-}
\end{eqnarray}

The corresponding Saha equations are
\begin{eqnarray}
    \frac{n_{\ce{He+}}n_{\ce{e-}}}{n_{\ce{He}}}&=&\frac{z_{\ce{He+}}z_{\ce{e-}}}{z_{\ce{He}}}=q_3  \label{eqn:hesaha1}\\
    \frac{n_{\ce{He^{2+}}}n_{\ce{e-}}}{n_{\ce{He+}}}&=&\frac{z_{\ce{He^{2+}}}z_{\ce{e-}}}{z_{\ce{He+}}}=q_4 \label{eqn:hesaha2}
\end{eqnarray}

The number conservation equation and charge neutrality equations are,
\begin{eqnarray}
    n_{\ce{He}}+n_{\ce{He+}}+n_{\ce{He^{2+}}}&=&Y\rho/m_{\ce{He}}=\bar{Y}.    \label{eqn:nheconserv}\\
    n_{\ce{He+}}+2n_{\ce{He^{2+}}}&=&n_{\ce{e-}}    \label{eqn:heneutral}
\end{eqnarray}
Substitute Equation \ref{eqn:hesaha1}, \ref{eqn:hesaha2} and \ref{eqn:heneutral} into \ref{eqn:nheconserv}, we get a cubic equation,
\begin{eqnarray}\label{eqn:hecubic}
    n_{\ce{He^{2+}}}^{3}(q_{3}-4q_{4})+n_{\ce{He^{2+}}}^{2}q_{1}\bar{Y}&-&n_{\ce{He^{2+}}}q_{3}q_{4}(q_{4}+2\bar{Y}) \notag\\
    &+&q_{3}q_{4}^{2}\bar{Y}=0
\end{eqnarray}
Solve Equation \ref{eqn:hecubic} gives us $n_{\ce{He^{2+}}}$, then $n_{\ce{He+}}$ and $n_{\ce{e-}}$ can be solved by Equation \ref{eqn:hesaha2} and \ref{eqn:heneutral}, and $n_{\ce{He}}$ is calculated from \ref{eqn:nheconserv}. Meanwhile, we write down the differential form of Equation \ref{eqn:hesaha1}, \ref{eqn:hesaha2}, \ref{eqn:nheconserv}, and \ref{eqn:heneutral} and calculate $(\partial n_{i}/\partial\rho)_{T}$ and $(\partial n_{i}/\partial\rho)_{\rho}$ of helium species.

\begin{eqnarray}
\begin{pmatrix}
    -1 & 1 & 0  &   1 \\
    0 & -1 & 1  &   1 \\
    n_{\ce{He}} & n_{\ce{He+}} & n_{\ce{He^{2+}}}  &   0   \\
    0 & n_{\ce{He+}} & n_{\ce{He^{2+}}}  &   -n_{\ce{e-}}
\end{pmatrix}
\begin{pmatrix}
    d\ln n_{\ce{He}}     \\
    d\ln n_{\ce{He+}}    \\
    d\ln n_{\ce{He2+}}   \\
    d\ln n_{\ce{e-}}
\end{pmatrix}
=\nonumber\\
\begin{pmatrix}
    d\ln q_{3}      \\
    d\ln q_{4}      \\
    \bar{Y}d\ln\bar{Y}         \\
    0
\end{pmatrix}
\end{eqnarray}
If only one reaction is present, the thermodynamic system is trivial to solve.

\begin{figure}[!ht]
    \centering
    \includegraphics[width=\columnwidth]{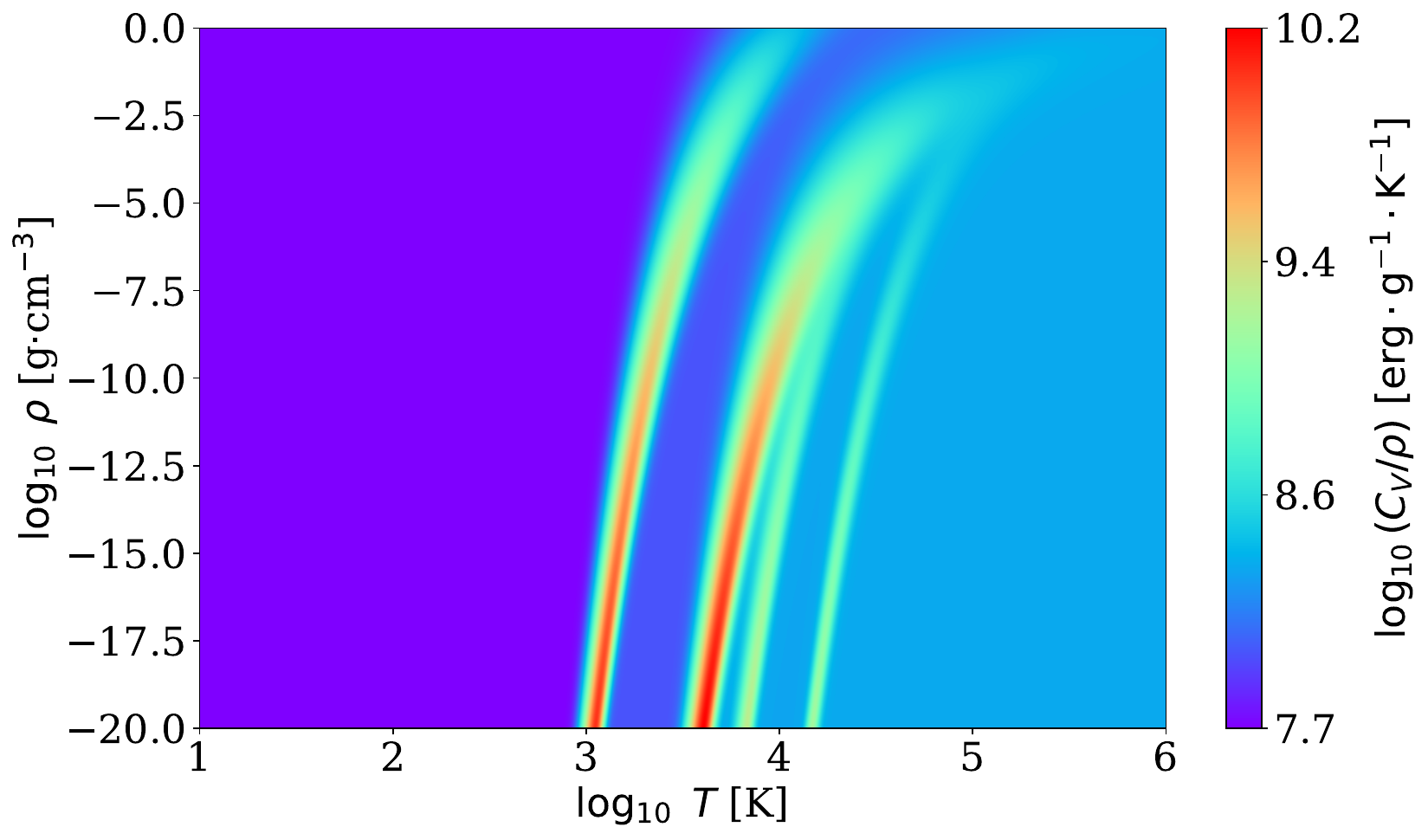}
    \includegraphics[width=\columnwidth]{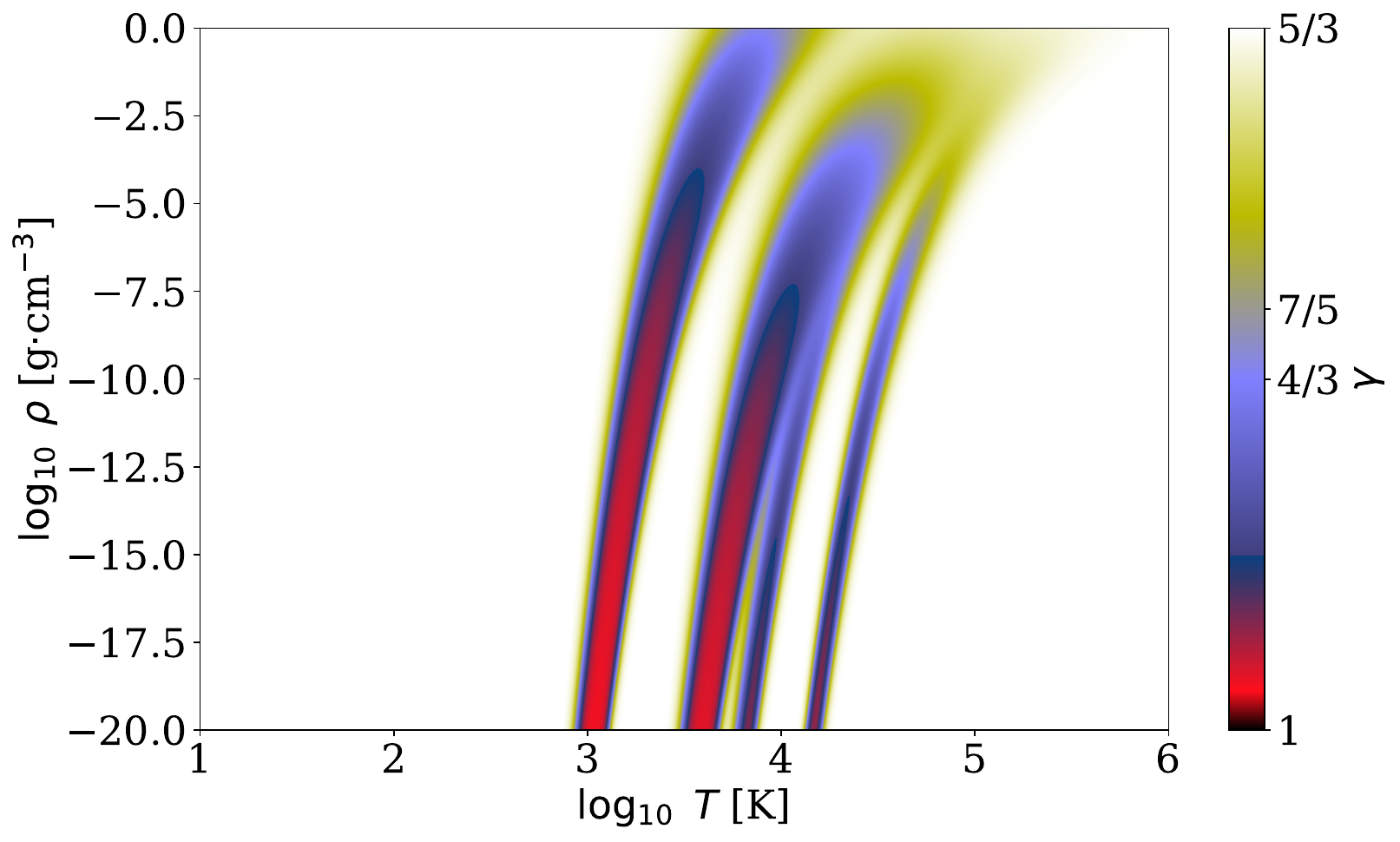}
    \caption{The first panel shows the combined specific heat capacity of $X=0.74$ and $Y=0.26$, and the bottom panel shows the adiabatic index.}
    \label{fig:gamma}
\end{figure}

Finally, we can calculate all the thermodynamic variables. Figure \ref{fig:gamma} shows the specific heat capacity and adiabatic index we use in this Letter.

\section{Opacity}\label{app:opacity}

\begin{figure}
    \centering
    \includegraphics[width=\columnwidth]{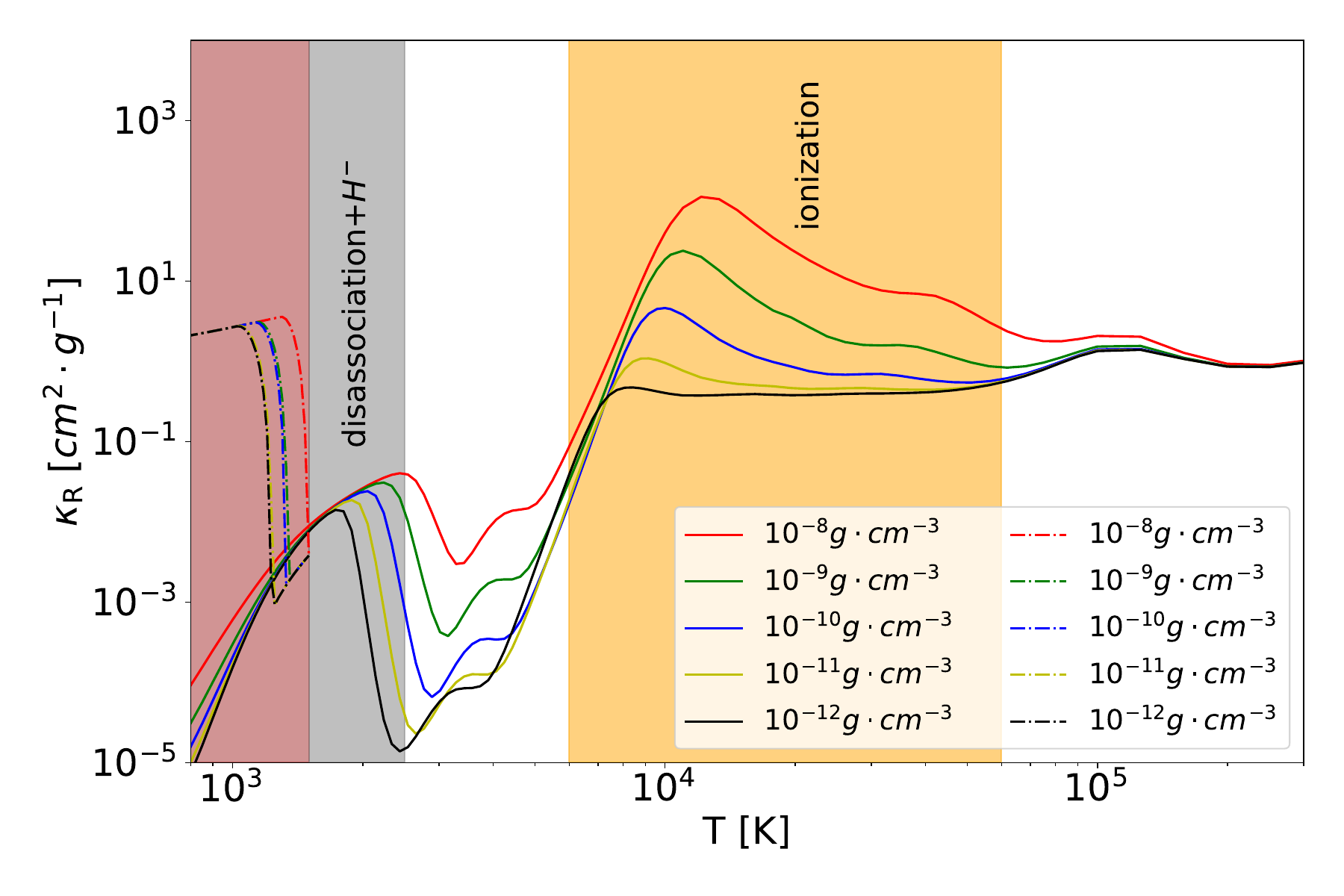}
    \caption{The Rosseland mean opacity that includes dust opacity at different densities. Solid and dashed lines denote the gas and dust opacities, respectively. The red zone indicates the dust formation temperature, the gray zone indicates the temperature of hydrogen dissociation and \ce{H-} formation, and the orange zone indicates the ionization temperature of \ce{H} and \ce{He}.}
    \label{fig:kr}
\end{figure}

The opacity table combines the gas opacity table of \citet{malygin2014} that dominates over $1500$K, and the dust opacity table (1\% of dust to gas ratio) of \cite{semenov2003} that dominates below 1100-1200K, depending on the density. For temperatures in between, the maximum of the value of the two tables is taken. Figure \ref{fig:kr} shows $\kr$ at different densities and $\kp$ is similar.

\section{time-dependent boundary conditions}\label{app:boundconds}

When $t<t_{1}$, the ejecta is radiation-dominated,
\begin{eqnarray}
    \dot{M}=\dot{M}_{1}+\frac{(\dot{M}_{2}-\dot{M}_{1})\exp{x}}{\exp{x}+1}+\dot{M}_{3}\bigg(\frac{t}{t_{1}}\bigg)^{2}
\end{eqnarray}
\begin{eqnarray}
    f_{\rm{ej}}=
    \begin{cases}
    1+\frac{(v_{1}-1)t}{x_{2}t_{1}}, \quad\quad\quad\quad\quad\quad\quad\quad t\le x_{2}t_{1}\\
    1+(v_{3}-1)\exp{-\big(\frac{x_{1}(t-x_{2}t_{1})}{t_{1}}\big)^{2}},\text{else}
    \end{cases}
\end{eqnarray}

When $t_{1}\le t\le t_{1}+t_{2}$, the ejecta becomes matter-dominated,
\begin{eqnarray}
    \dot{M}&=&\dot{M}_{2}-\frac{(\dot{M}_{2}-\dot{M}_{1})\exp{-x}}{\exp{-x}+1}-\delta\dot{M}\frac{t-t_1}{t_2}  \\
    f_{\rm{ej}}&=&v_{2}+\big(1+(v_{3}-1)\exp{-(x_{1}(1-x_{2}))^{2}}-v_{2}\big)\nonumber\\ 
    &&\exp{\frac{t-t_{1}}{dt_{2}}}+\delta v\big(\frac{t-t_{1}}{t_{2}}\big),
\end{eqnarray}
where
\begin{eqnarray}
    v_{\rm{esc}}&=&\sqrt{2G\M/r_{\rm{in}}}, \\
    \delta\dot{M}&=&(\dot{M}_{2}-\dot{M}_{1}), \\
    x&=&20(t-t_{1})/t_{1}.
\end{eqnarray}
There are 13 parameters in this model, we list all the parameters just to show the functional form of the ejecta. In terms of the shock and shock-free models, they have many parameters in common. The common parameters are: $t_{1}=1.3$ days, $dt_{2}=1.6$ days, $v_{1}=1.06$, $\dot{M}_{1}=0.16$\msunyr, $\dot{M}_{3}=0.2$\msunyr, $x_{1}=6$, and $x_{2}=0.55$. We list the different parameters in Table\ref{tab:parameters}.
\begin{table}
    \centering
    \begin{tabular}{cccccccc}\hline
      &   $t2$  &   $v2$  &   $v3$    &   $\delta v$  &   $\dot{M}_2$  &   $\delta\dot{M}$  \\
       &   [day]   &   &   &   &   [\msunyr]    &   [\msunyr]   \\\hline   
    s  &   8   &   0.87    &   0.94    &   0.01     &   1.3 &   0   \\
    sf &   20  &   0.875   &   0.955   &   -0.04    &   1.8 &   -0.5    \\\hline
    \end{tabular}
    \caption{The parameters that are different in shock (s) and shock-free (sf) models. In particular, $\delta v>0$ cases generally belong to shock models and $\delta v<0$ cases belong to shock-free models.}
    \label{tab:parameters}
\end{table}

We anticipate that there could be other fitting functions with more/less parameters. One should explore different functional forms and a thorough parameter space study is deemed appropriate in the future.

\section{The time evolution of hydrogen and helium species}\label{app:hhe}

\begin{figure}
    \centering
    \includegraphics[width=\columnwidth]{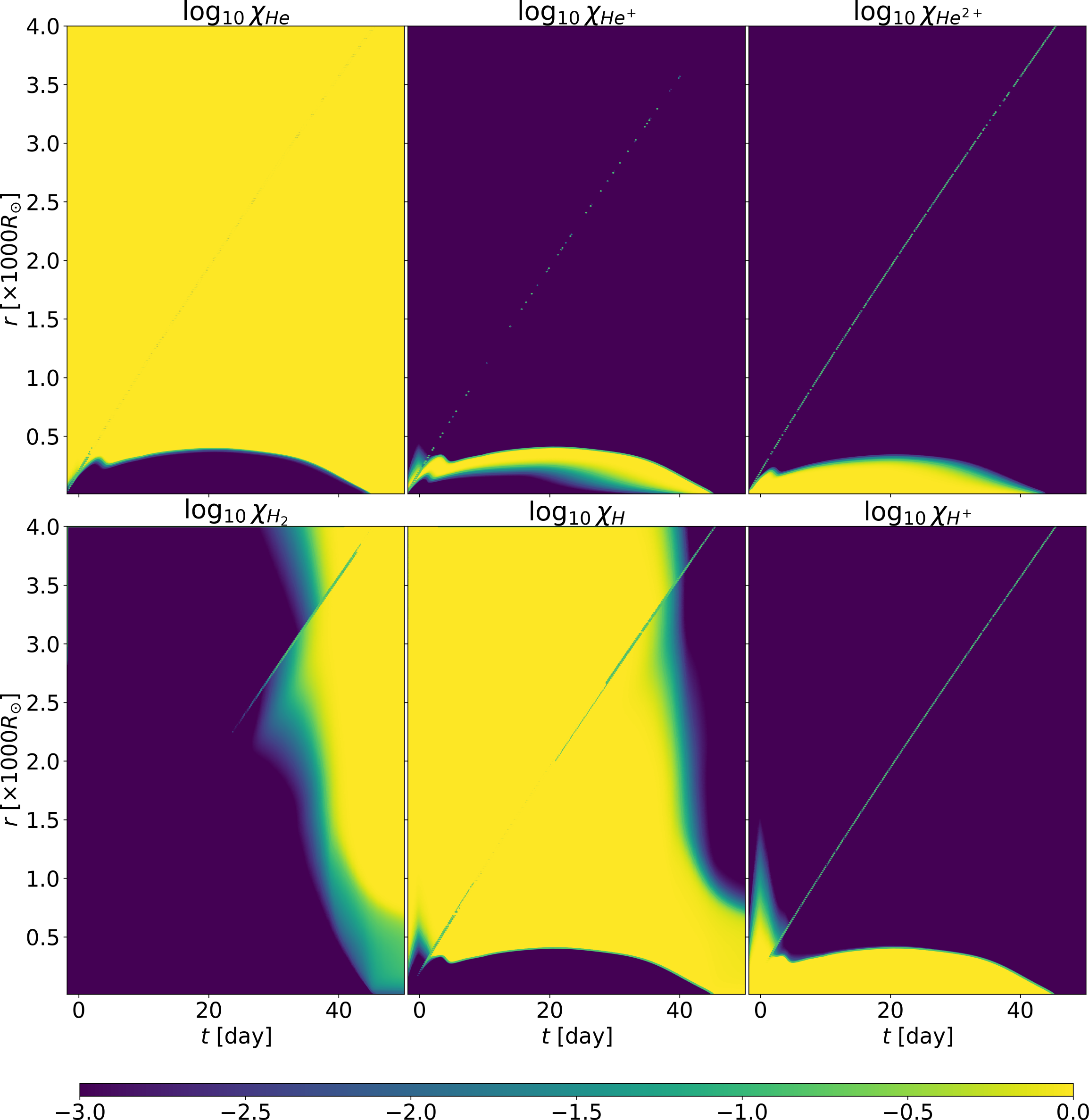}
    \caption{From the left to the right, the first row shows the time evolution of the mass fraction of \ce{He}, \ce{He+}, and \ce{He^{2+}}, the second row shows the time evolution of the mass fraction of \ce{H2}, \ce{H}, and \ce{H+}.}
    \label{fig:kippenhahnhhe}
\end{figure}

Figure \ref{fig:kippenhahnhhe} shows the time evolution of the mass fraction of \ce{H} and \ce{He} species of the shock model. For example, the mass fraction of \ce{H2} is defined as,
\begin{equation}
    \chi_{\ce{H2}}=\frac{2n_{\ce{H2}}}{2n_{\ce{H2}}+n_{\ce{H}}+n_{\ce{H+}}}
\end{equation}

Although the mass fraction of each species can be derived from $\rho$ and $\tg$ already, there are two features we would like to clarify.
\begin{enumerate}[1.]
    \item \ce{H2} forms in the late stage. There are two channels for \ce{H2} formations. At high density, \ce{H} can form \ce{H2} directly \citep{omukai2005}. When dust is present, \ce{H2} can form on dust \citep{wakelam2017}. In the time evolution, we find that the \ce{H2} formation region largely coincide with the dust formation region (can be seen from the $\log_{10}\kr$ panel in Figure \ref{fig:timeevolmisc}) or with the high-density region (can be seen from the $\log_{10}\rho$ panel in Figure \ref{fig:timeevolmisc}). Therefore, Saha's EoS model is suitable for this problem.
    \item There is a line showing the partial ionization of \ce{H+} in the \ce{H+} panel. This is due to Zeldovich's spikes in the radiation hydrodynamic calculation after a shock. The shock is caused by the fast ejecta colliding with the ambient. However, we tend not to ascribe the H$\alpha$ emission to this line because the decreasing width of H$\alpha$ in observations \citep{pastorello2021a} indicates that the shock decelerates. We believe that the H$\alpha$ emission is caused by a marginally escaping ejecta colliding with the pre-existing decreation disk \citep{pejcha2017,metzger2017}. However, the 1D model is only shell-like and cannot model disks.
\end{enumerate}

\section{The impact of realistic EoS}\label{sub:impacteos}

\begin{figure}
    \centering
    \includegraphics[width=\columnwidth]{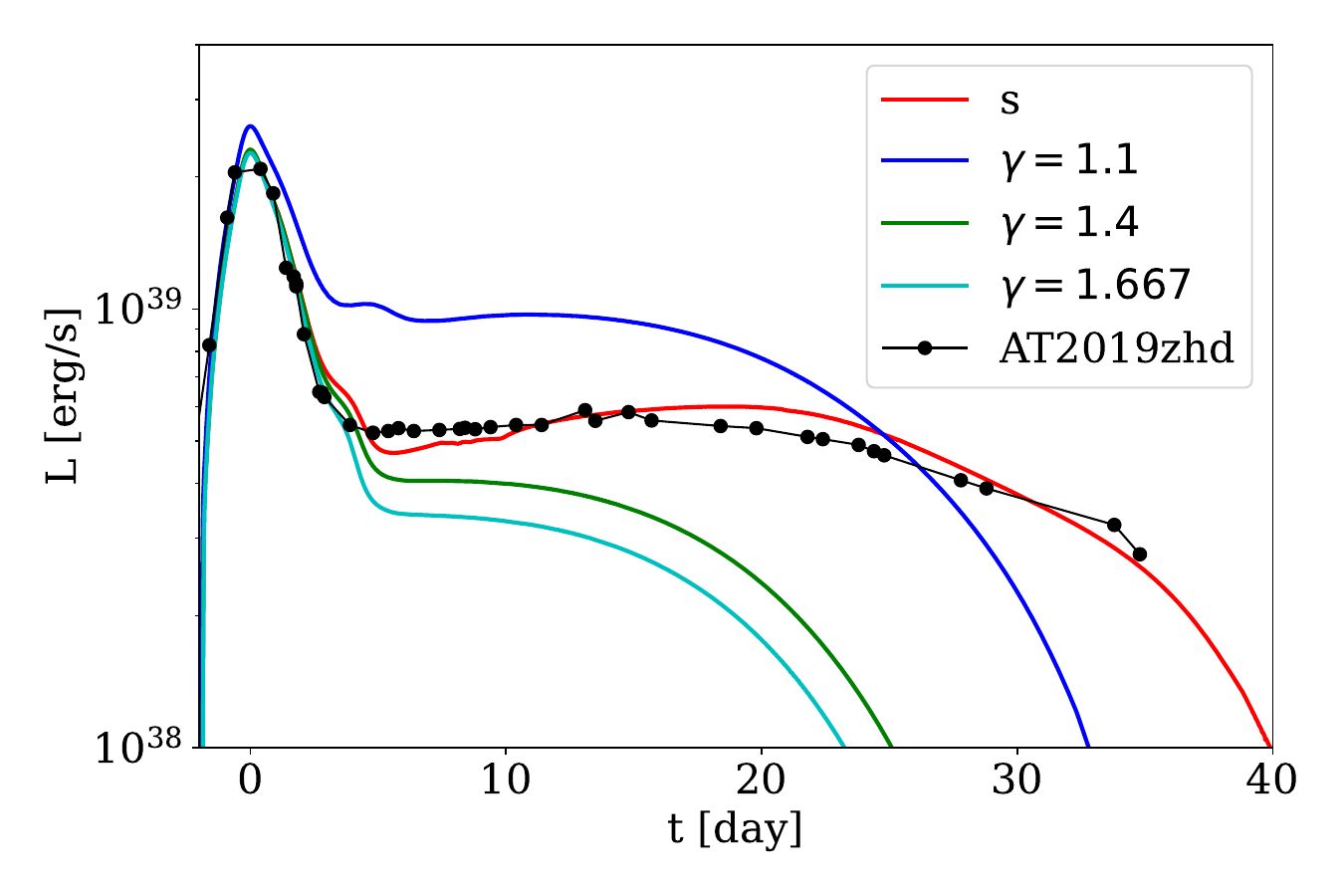}
    \includegraphics[width=\columnwidth]{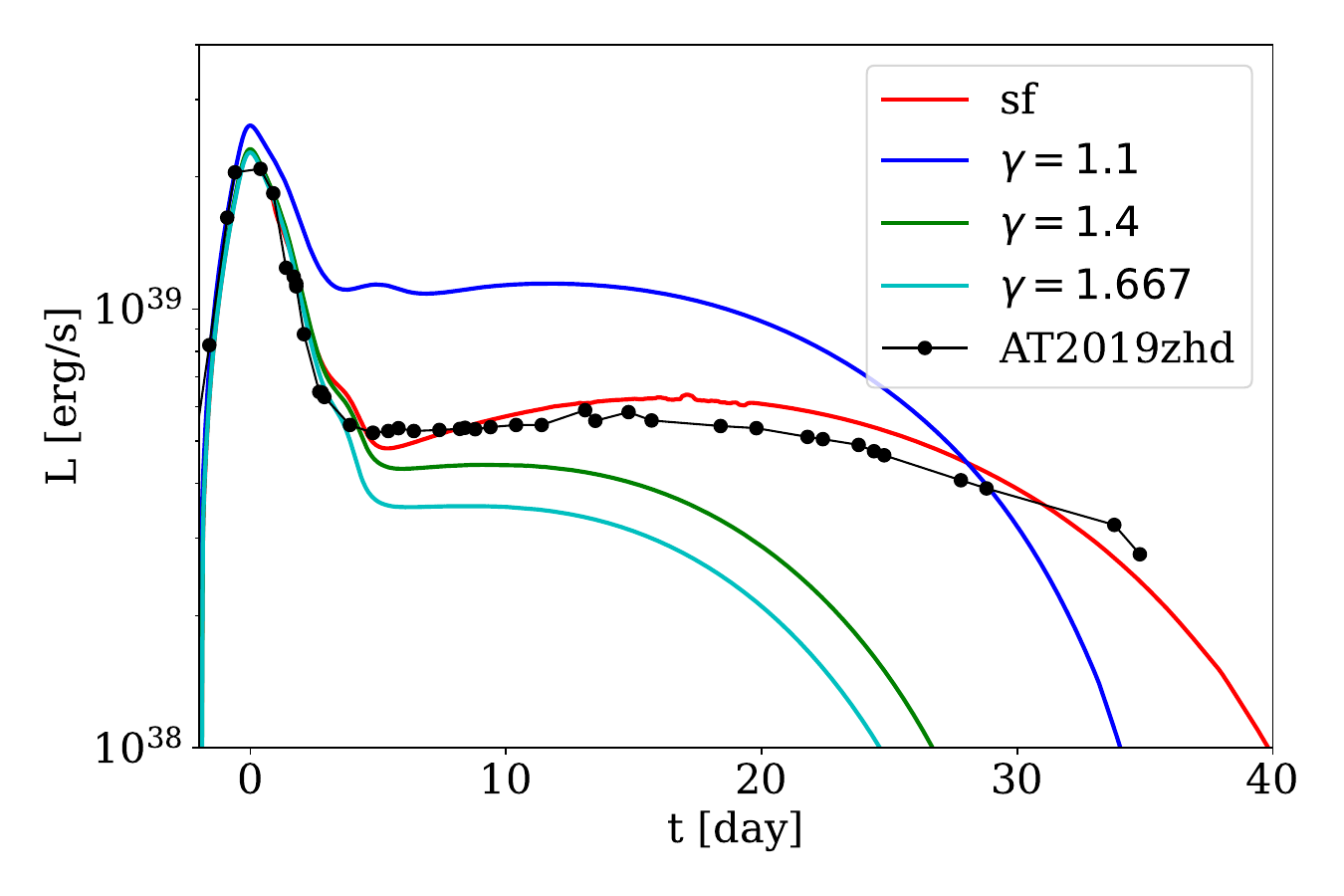}
    \caption{The upper and lower panels show the light curves of \ce{H} and \ce{He} mixture EoS and perfect gas EoS with various $\gamma$ of the shock (s) and shock-free (sf) models, respectively.}
    \label{fig:lightcurvegamma}
\end{figure}

EoS can play an important role in radiation hydrodynamics. 
We compare our simulations with realistic EoS to the widely used in hydrodynamic codes $\gamma$-law EoS, described by
\begin{eqnarray}
    \eg&=&\rho k_{b}\tg/(\gamma-1),     \\
    p&=&\rho k_{b}\tg/(\mu m_{H}),
\end{eqnarray}
where $\mu=1$ is the mean molecular weight.

With $\gamma$-law EoS, gas energy does not contain latent heat. As a result, the cooling timescale is shorter, and all the light curves ran with $\gamma$-law EoS last for a shorter time. We can also see that the peak is less affected by the change in the EoS because the radiation-dominated ejecta governs the peak. The internal energy $\eg$ plays a minor role in this stage.

\section{The impact of radiation force}\label{sub:frad}

\begin{figure}
    \centering
    \includegraphics[width=\columnwidth]{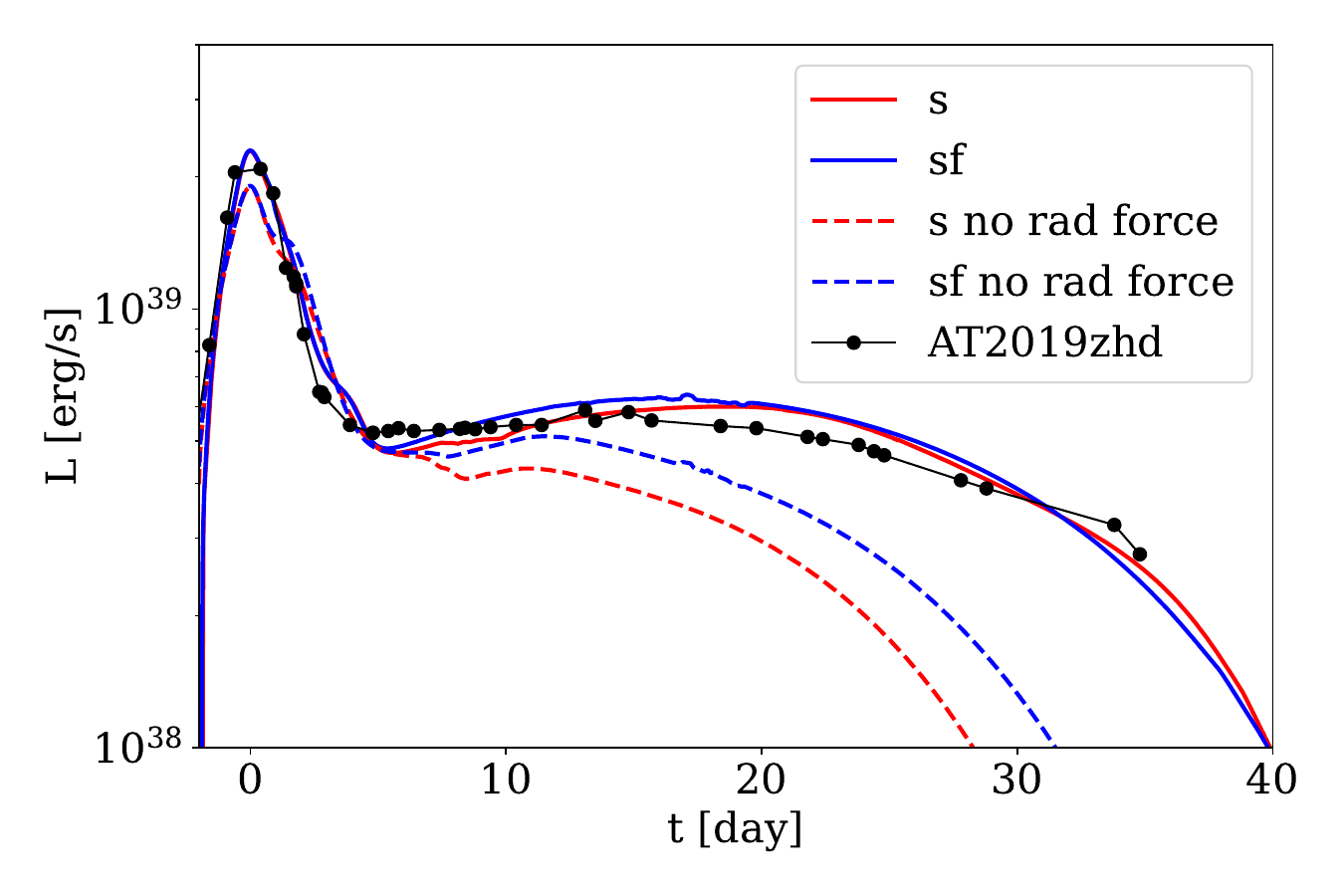}
    \caption{The black line with dots shows the light curve of AT2019zhd. The red and blue colors indicate the shock (s) and shock-free (sf) models. The solid and dashed lines indicate our original models and their counterparts when we turn off the radiation force.}
    \label{fig:nofrad}
\end{figure}

Radiation force was largely ignored in previous hydrodynamical simulations of CEE. However, we find that the $a_{\rm{rad}}$ is larger than $g$ at high ($>10000$K) and low temperatures ($<1400$K); see panel 3 in Figure \ref{fig:timeevolmisc}. To demonstrate the importance of radiation force, we compare our original simulations' light curves to those without radiation force ($a_{\rm{rad}}=0$). The results are presented in Figure \ref{fig:nofrad}.

In the case of no radiation force in the momentum equation, the luminosity drops off faster. This faster drop is due to more material in the ejecta falling back to the central object under the influence of the gravitational force. The drop happens so early that the gas is still fully ionized and should be pushed away by the radiation force. In our model, the fallback material will pass the inner boundary and no longer contribute to the luminosity.

\section{The impact of dust formation}\label{sub:dust}

\begin{figure}[!ht]
    \centering
    \includegraphics[width=\columnwidth]{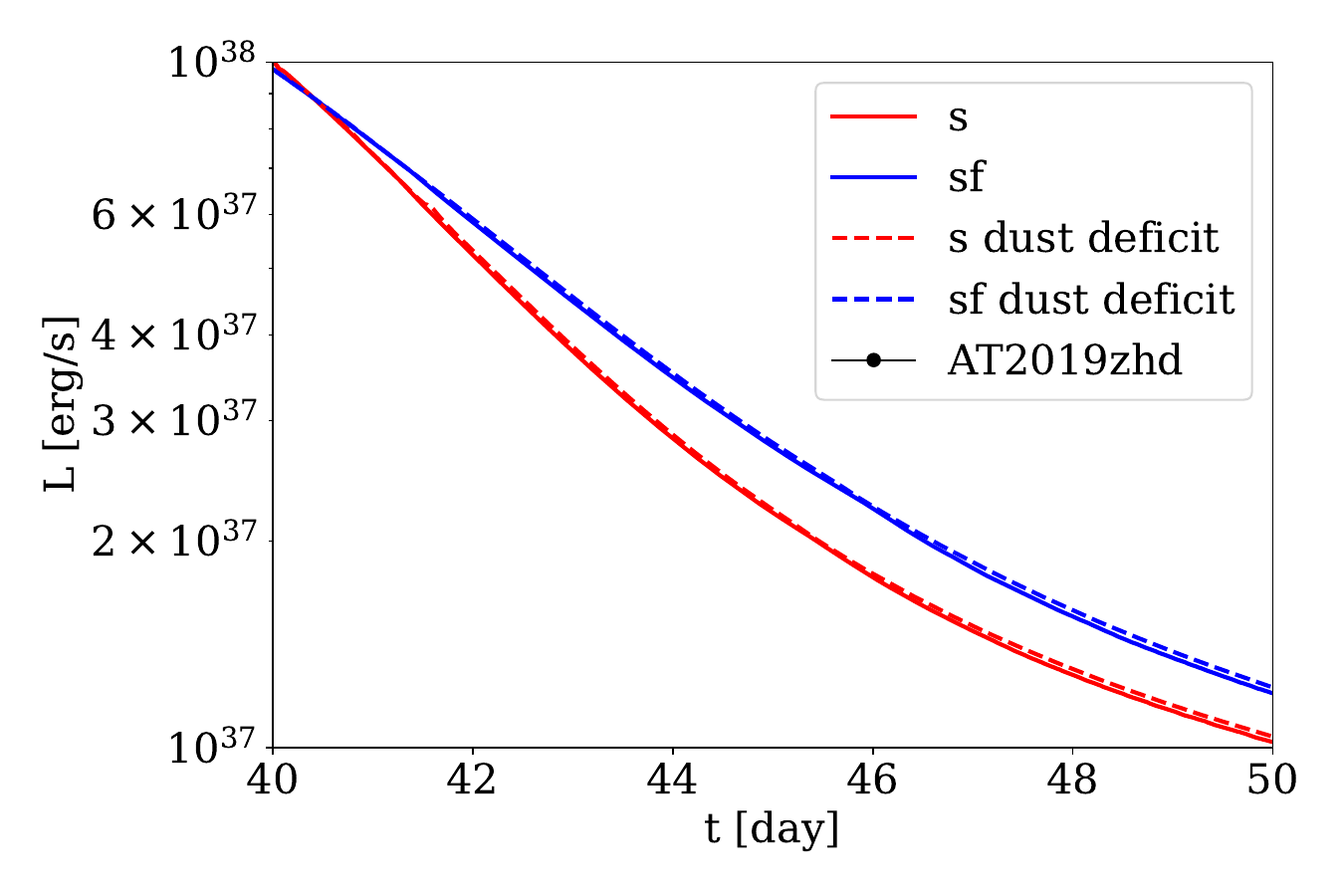}
    \includegraphics[width=\columnwidth]{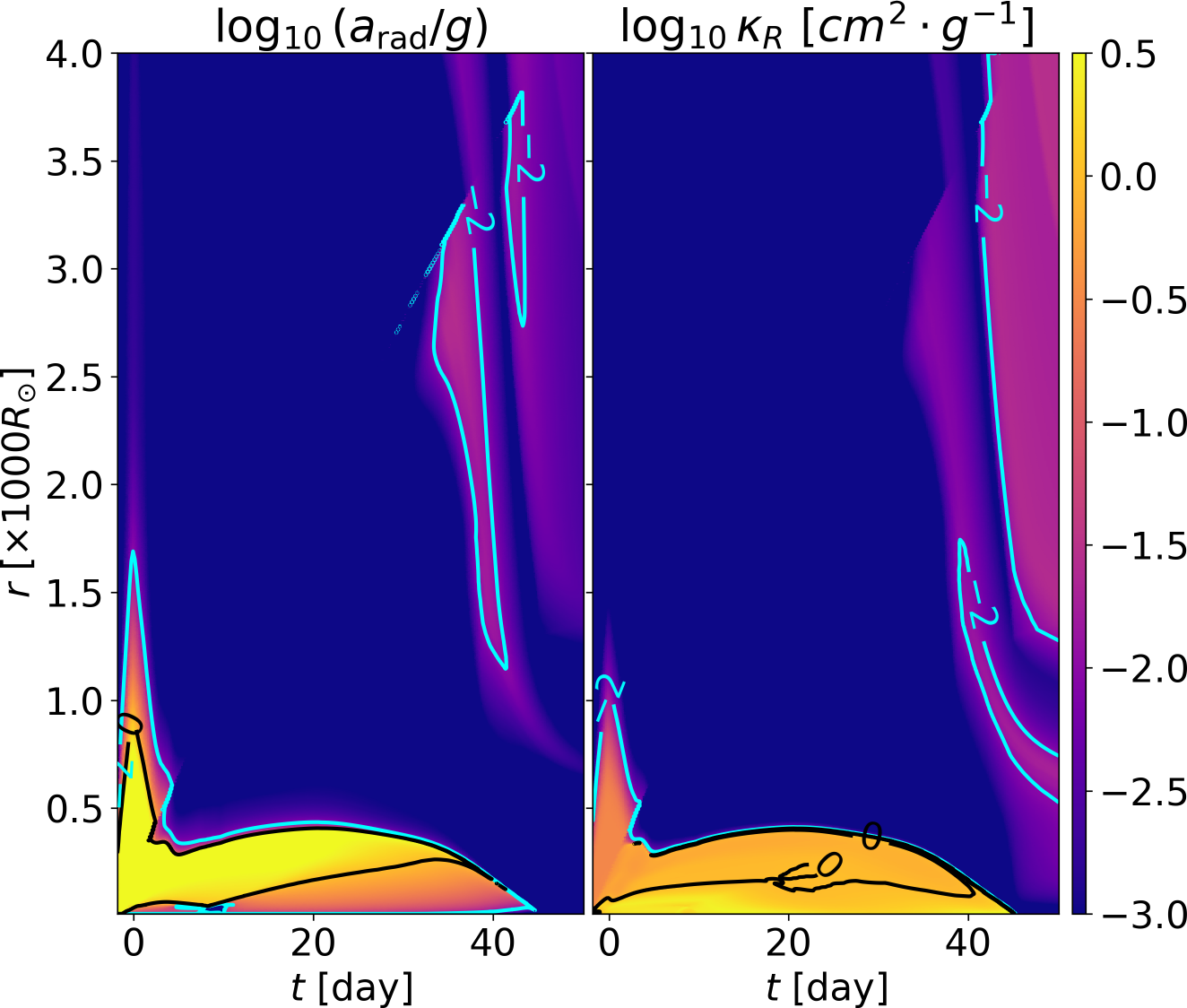}
    \caption{Upper panel: light curves of the last 10 days of shock (s) and shock-free (sf) models and their corresponding dust deficit models. Bottom panel: the time evolution of $\log_{10}(\kr\F/(cg))$ and $\log_{10}\kr$ of the shock dust deficit model.}
    \label{fig:nodust}
\end{figure}

It is argued that molecules and dust form during the late stage of CEE \citep{nicholls2013,kaminski2015,banerjee2015,blagorodnova2020,iaconi2020,morgan2022}. Numerical models also show that dust may provide an additional push away from the binary merger due to radiation acceleration \citep{gonzalez2023}. To test the impact of dust formation, we modify the opacity table (Figure \ref{fig:kr}) by decreasing $\kr$ and $\kp$ of dust by 100 times and name it the dust deficit model. We find that modification of dust opacities does now affect light curves during the peak and the plateau. This is unsurprising, as the temperature is too high during the early stage of LRNe for dust to form. The minor difference is that dust-deficit models have slightly higher luminosity during the very late stage (see Figure \ref{fig:nodust}), mainly because the optical depth of the dust region is smaller. 

Figure \ref{fig:nodust} also shows the time evolution of $\log_{10}(a_{\rm{rad}}/g)$ and $\log_{10}\kr$ of the dust-deficit model. Comparing Figure \ref{fig:nodust} to Figure \ref{fig:timeevolmisc}, we notice that the difference is mainly in the upper right corner of the figure, caused by the lowered dust opacity. The time and radial location of the dust formation is not changed. It indicates that the dust formation may have a limited impact on the radiation hydrodynamics of LRNe events during the early stage. In the late stage, due to radiation force acceleration, dust formation may play a role in shaping the ejecta, but still not a major physical mechanism of unbinding the ejecta \citep{bermudez2024}.

\section{convergence}\label{app:convergence}

\begin{figure}
    \centering
    \includegraphics[width=\columnwidth]{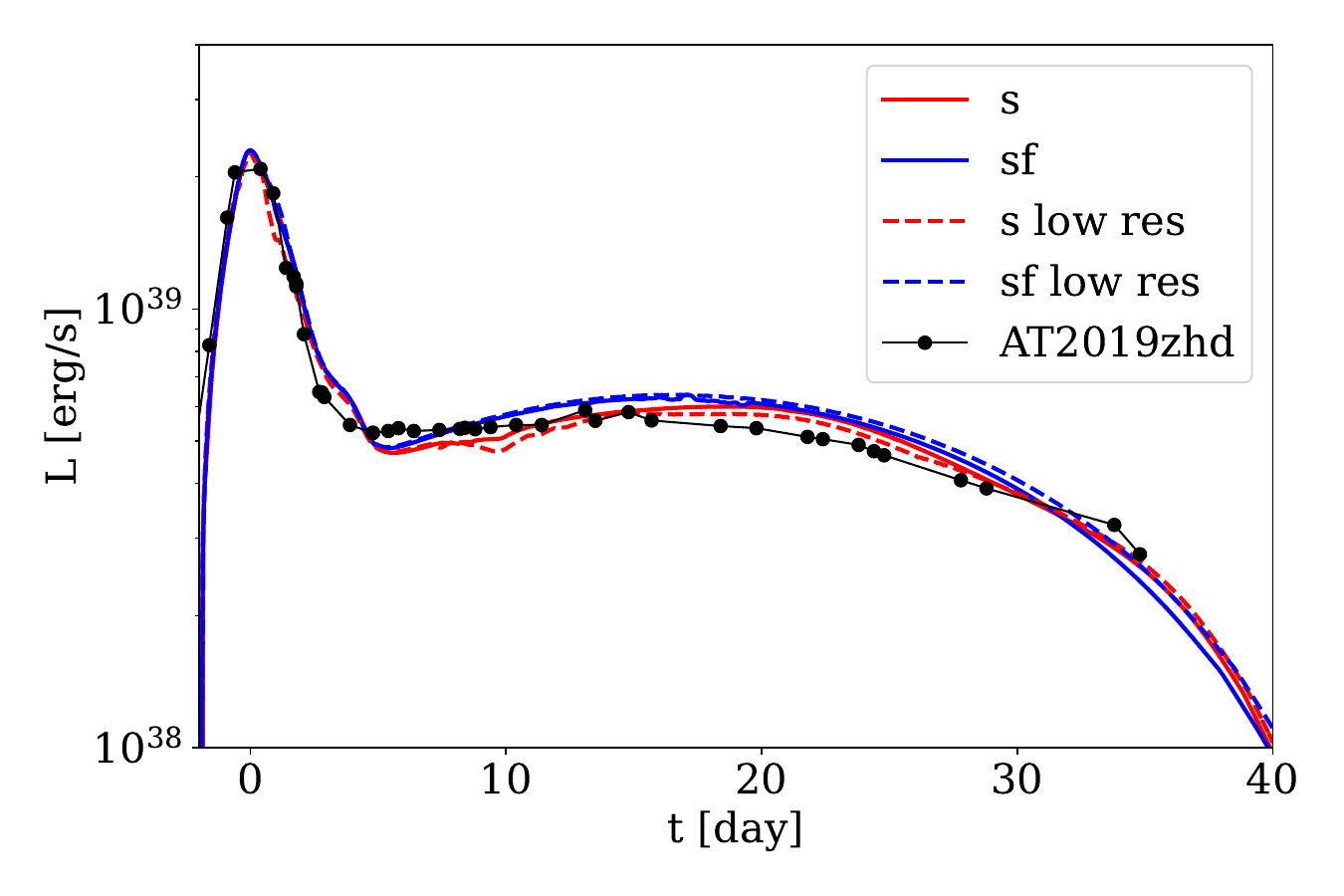}
    \caption{The black line with dots shows the light curve of AT2019zhd. The red and blue colors indicate the shock (s) and shock-free (sf) models. The solid and dashed lines indicate the original and low-resolution models, respectively.}
    \label{fig:convergence}
\end{figure}

Convergence is very important for simulations that have iterative solvers and mesh refinement. It is also well-known that radiation hydrodynamic equations may be sensitive to the choice of timesteps. We show that our simulations approach convergence by comparing the light curves of our original resolution to a lower base resolution ($N=768$), which also doubles the timesteps. Figure \ref{fig:convergence} has the light curves of the high and low resolutions of the shock and shock-free models. We can see that they are similar, indicating that our results are close to convergence.

\end{CJK*}
\end{document}